\documentclass[fleqn,usenatbib]{mnras}

\usepackage{newtxtext,newtxmath}
\usepackage{xcolor} % for text coloring
\usepackage[T1]{fontenc}

\DeclareRobustCommand{\VAN}[3]{#2}
\let\VANthebibliography\thebibliography
\def\thebibliography{\DeclareRobustCommand{\VAN}[3]{##3}\VANthebibliography}

\newcommand{\radms}{\text{rad\,m$^{-2}$}}

\usepackage{graphicx}	% Including figure files
\usepackage{amsmath}	% Advanced maths commands

\usepackage{pdflscape}
\usepackage{threeparttable}

\title[RM Periodicity in a Helical Jet]{Helical radio jets as probes of magnetised cluster environments: Periodic Faraday Rotation Revealed in the Corkscrew Galaxy by POSSUM}

\author[S. N. Bradbury et al.]
{Sarah\ N.\ Bradbury$^{1,2}$\thanks{E-mail: sarah.bradbury@anu.edu.au},
C.\ S.\ Anderson$^{1}$,
N.\ M.\ McClure-Griffiths$^{1,3}$,
Y.\ K.\ Ma$^{1,4}$,
A.\ J.\ M.\ Thomson$^{5,2}$,
\newauthor
L.\ Rudnick$^{6}$,
S.\ P.\ O'Sullivan$^{7}$,
B.\ M.\ Gaensler$^{8,9,10}$,
B.\ S.\ Koribalski$^{11,12}$,
E.\ L.\ Alexander$^{13}$,
\newauthor
Takuya Akahori$^{14}$,
L.\ Baidoo$^{9}$,
E.\ Carretti$^{15}$,
G.\ Heald$^{5,2}$,
H.\ Sakemi$^{16,17}$,
T.\ Vernstrom$^{2}$
and J.\ West$^{18,19}$
\\
$^{1}$Research School of Astronomy \& Astrophysics, Australian National University, Canberra, ACT 2611, Australia\\
$^{2}$ATNF, CSIRO Space \& Astronomy, Bentley, WA, Australia\\
$^{3}$SKA Observatory, Jodrell Bank, Lower Withington, Macclesfield, SK11 9FT, UK\\
$^{4}$Max-Planck-Institut f\"ur Radioastronomie, Auf dem H\"ugel 69, 53121 Bonn, Germany\\
$^{5}$SKA Observatory, SKA-Low Science Operations Centre, 26 Dick Perry Avenue, Kensington WA 6151, Australia\\
$^{6}$University of Minnesota, Minneapolis, MN 55455, USA\\
$^{7}$Departamento de Física de la Tierra y Astrofísica \& IPARCOS-UCM, Universidad Complutense de Madrid, 28040 Madrid, Spain\\
$^{8}$Department of Astronomy and Astrophysics, University of California Santa Cruz, 1156 High Street, Santa Cruz, CA 95064, USA\\
$^{9}$Dunlap Institute for Astronomy and Astrophysics, University of Toronto, 50 St. George St, Toronto, ON M5S 3H4, Canada\\
$^{10}$David A. Dunlap Department of Astronomy and Astrophysics, University of Toronto, 50 St. George St, Toronto, ON M5S 3H4, Canada\\
$^{11}$Australia Telescope National Facility, CSIRO, Space and Astronomy, P.O. Box 76, Epping, NSW 1710, Australia\\
$^{12}$Western Sydney University, Locked Bag 1797, Penrith South DC, NSW 2751, Australia\\
$^{13}$School of Physics \& Astronomy, University of Leeds, Leeds, LS2 9JT, UK\\
$^{14}$Mizusawa VLBI Observatory, National Astronomical Observatory of Japan, 2-21-1 Osawa, Mitaka, Tokyo 181-8588, Japan\\
$^{15}$Istituto Nazionale di Astrofisica, Via Gobetti 101, 40129 Bologna, Italy\\
$^{16}$Graduate School of Sciences and Technology for Innovation, Yamaguchi University, 1677-1 Yoshida, Yamaguchi, 753-0841, Japan\\
$^{17}$ Nobeyama Radio Observatory, National Astronomical Observatory of Japan (NAOJ), National Institutes of Natural Sciences (NINS), 462-2, Nobeyama,\\ Minamimaki, Minamisaku, Nagano 384-1305, Japan\\
$^{18}$Dominion Radio Astrophysical Observatory, Herzberg, National Research Council, Penticton, BC V2A 6J9, Canada\\
$^{19}$School of Natural Sciences, University of Tasmania, PO Box 807, Sandy Bay, TAS 7006, Australia}

\date{Accepted XXX. Received YYY; in original form ZZZ}

\pubyear{\the\year{}}

\begin{document}
\label{firstpage}
\pagerange{\pageref{firstpage}--\pageref{lastpage}}
\maketitle


\begin{abstract}
Jets from active galactic nuclei (AGN) that exhibit regular plane-of-sky oscillations may provide a probe of magnetised plasma within the jets and their local intracluster medium (ICM). If such morphologies trace three-dimensional helices, path length differences between near and far sides of the flow might imprint quasi-periodic signatures in Faraday rotation. We present ASKAP Band~2 (1296–1440 MHz) polarimetric observations of a helical-tailed radio galaxy (the ``Corkscrew Galaxy") from early science data from the POlarisation Sky Survey of the Universe’s Magnetism (POSSUM), and test whether its quasi-periodic morphology produces signatures in Faraday rotation. We detect significant RM oscillations with a spatial period of $(342 \pm 101)$ arcsec, consistent within the Rayleigh resolution limit with the jet’s lateral deviations ($(290 \pm 72)$ arcsec). Cross-correlation analysis reveals systematic variation along the jet: the eastern section shows alignment between jet deviation and RM, consistent with a jet-associated or sheath-like Faraday screen, while the western section exhibits a phase shift, indicating a transition in the dominant Faraday-rotating medium and possible contribution from the local ICM. These results demonstrate that quasi-periodic RM signatures can disentangle the dominant Faraday-rotating medium in AGN jets, and show that in some embedded radio galaxies the dominant RM contribution arises near the source rather than the cluster foreground. We also outline reliability criteria required to avoid false positive detections and inadequate sampling. Helical jet structures therefore show strong promise as probes of magnetised cluster environments, a capability that will expand in the SKA era.
\end{abstract}

\begin{keywords}
galaxies: clusters: individual – galaxies: clusters: intracluster medium – magnetic fields – radio continuum: galaxies –
techniques: polarimetric
\end{keywords}

%%%%%%%%%%%%%%%%%%%%%%%%%%%%%%%%%%%%%%%%%%%%%%%%%%
    
%%%%%%%%%%%%%%%%% BODY OF PAPER %%%%%%%%%%%%%%%%%%

\section{Introduction}\label{sec:intro}
Radio galaxies host active galactic nuclei (AGN) that produce intense radio emission from relativistic jets and lobes, often extending well beyond their optical host galaxies \citep{Miley2008}. These jets, launched by accretion onto supermassive black holes, can propagate hundreds of kiloparsecs into the intergalactic medium, driving shocks, inflating lobes, and redistributing energy and magnetic fields on large scales \citep{Bridle1984, Begelman1984}. Through these processes, radio galaxies play a central role in regulating the growth of galaxies and their environments \citep{Fabian2012}.

Magnetic fields are critical to the physics of these systems: they help launch and collimate jets, and govern their interaction with local plasma. Observationally, Faraday rotation—the wavelength-dependent rotation of the polarisation angle of linearly polarised radiation as it passes through magnetised plasma—provides a powerful diagnostic of these fields \citep{Burn1966, Govoni2004}. By mapping the distribution of rotation measures (RMs) across a radio source, one can test whether the observed Faraday rotation primarily occurs in the jet itself, in a sheath of entrained plasma, or in the foreground intracluster medium (ICM). Distinguishing between these scenarios is a central challenge for understanding how AGN jets trace, and are shaped by, their environments (e.g. \citealp{Anderson2018}). Hereafter, we use the following terms to distinguish between different locations of Faraday rotating media: `internal' to mean the Faraday rotating medium inside the synchrotron emitting region, `local' to mean the Faraday rotating medium in either a sheath-like structure around the radio jets, or within the ICM immediately surrounding the jets (but within the cylindrical volume enclosed by the helical jets), and `foreground' to mean either Faraday rotation occurring in the Milky Way, or in the ICM in the foreground of the Corkscrew (outside the cylindrical volume enclosed by the helical jets). 

Helical and oscillatory structures are observed across a wide range of radio jet systems, from parsec to megaparsec scales, when sufficient 
angular resolution is available \citep[e.g. M87;][]{Pasetto2021}. On larger scales, a subset of tailed radio galaxies exhibit highly extended ($\gtrsim 100$~kpc), often one-sided jets with pronounced helical substructure, including IC~711 in Abell~1314 and the so-called ``MysTail'' in Abell~3266 \citep{Condon2021, Rudnick2021}, as well as sources recently identified in the Ophiuchus cluster \citep{Botteon2025}. Whether such helical structure is intrinsically rare or simply undetected in most tailed sources due to insufficient angular resolution or sensitivity remains an open question. The increasing resolution and depth of surveys with ASKAP and MeerKAT are beginning to reveal that wiggles and transverse substructure may be more common in extended radio galaxies in general, than previously realised \citep{Koribalski2024, Botteon2025}.

Corkscrew-shaped radio galaxies provide a promising opportunity to investigate the connection between jet morphology and Faraday rotation. Their repeating, helical structures introduce cyclic variations in line-of-sight (LOS) depth through both the jet and its local medium, enabling tests of whether RM variations are linked to local jet physics or to the gaseous environment in which the source is embedded. If the RMs reflect the internal magnetohydrodynamic state of the jet, they may reveal the influence of instabilities such as Kelvin–Helmholtz or kink modes in shaping its structure \citep[e.g.][]{Kigure2004, Nakamura2007, Mizuno2014, Singh2016, BarniolDuran2017, Upreti2024}. Conversely, if the RMs arise primarily from magnetised intracluster gas enclosed by the helices, corkscrew sources could serve as powerful in-situ probes of cluster magnetisation \citep{Pratley2013, Johnston-Hollitt20151}. If instead the RMs vary independently of the corkscrew morphology, this would suggest that they mainly probe the integrated magnetised plasma along the LOS, whether in the ICM or the Milky Way foreground. Thus, corkscrew radio galaxies offer a valuable new avenue for exploring two closely related problems in contemporary astrophysics: the physics and stability of AGN jets, and the magnetic field structure of galaxy clusters.

Until recently, polarimetric studies of helical jet structures were largely confined to parsec-scale systems observed with very long baseline interferometry (e.g. 3C~345, M87, 3C~120; \citealt{Zensus1995, Steffen1995, Reid1989, Owen1989}). Advances in wide-field radio facilities such as ASKAP and MeerKAT now allow systematic studies of sources with similar morphologies on much larger (tens to hundreds of kiloparsec) scales \citep{Condon2021,Koribalski2024}. The Polarisation Sky Survey of the Universe’s Magnetism \citep[POSSUM;][]{Gaensler2025} on ASKAP, in particular, is designed to combine broad frequency coverage, high sensitivity, and excellent polarisation fidelity, enabling new tests of whether corkscrew radio galaxies can serve as reliable probes of Faraday rotation in cluster environments.

ESO~137$-$G007—the “Corkscrew Galaxy” (1610$-$60.5)—is the most striking known example of this class. Located near the periphery of the Norma cluster (Abell~3627; $z \simeq 0.016$), it exhibits a highly collimated, twisting radio tail extending $\sim$28~arcmin ($\sim$570~kpc) from the host galaxy \citep{Ekers1969, Jones_McAdam_1994, Jones_McAdam_1996}. Recent ASKAP 944~MHz (Band 1) and 1.4~GHz (Band 2) observations presented by \citet{Koribalski2024} reveal intricate filamentary and thread-like structures along the older portions of the tail, which they attribute to interactions with the ICM, including shear forces and the stripping of an outer cocoon. The radio peak is offset from the optical nucleus, indicating early jet bending prior to tail formation. Morphologically, the jet can be divided into two main regions separated near RA, Dec (J2000) $\sim16^{\mathrm h}13.5^{\mathrm m}$, $-60\degr34\arcmin$, where the flow appears disrupted, bends, and subsequently continues at the bent orientation. East of this point, the emission transitions from a knotty inner structure near the host galaxy to well-resolved helices further downstream, while to the west the jet develops broader, higher-amplitude corkscrew windings spanning approximately two full turns before dispersing into faint synchrotron filaments. The overall morphology is consistent with a dynamically perturbed or precessing jet whose evolution has been shaped by interaction with the local ICM \citep[see][for a detailed characterisation of the thermodynamic structure of the Norma cluster ICM]{Ge2026}.

In this work, we focus exclusively on ASKAP 1.4~GHz polarimetric observations (Band 2; 1296--1440~MHz) obtained during early POSSUM science operations. At these frequencies, the Corkscrew Galaxy exhibits substantially stronger and more spatially coherent linear polarisation than at lower frequencies, making the Band~2 data particularly well-suited to spatially resolved RM analysis along the jet. The ASKAP 944~MHz Band~1 results of \citet{Koribalski2024} are therefore referenced here to provide morphological and physical context, but are not re-analysed in this study.

With these POSSUM Band~2 polarimetric observations, our aim is to address a set of interlocking scientific and technical questions. Chief among the technical questions is whether the quasi-periodic jet morphologies of corkscrew-type sources show corresponding quasi-periodic variations in Faraday RM, and if so, whether such signatures trace magnetic fields internal to the jet, the local ICM, or unrelated foreground material, whether in the ICM or elsewhere along the LOS. From this follow a series of methodological considerations: how RM–morphology correlations can be quantified robustly, whether the Galactic foreground fluctuations or structured noise effects can mimic periodic RM signals, and what combination of angular resolution, sensitivity, and number of resolved windings is required for secure detection.

The outcomes of these technical tests will determine which broader science questions corkscrew radio galaxies can address. How are helical jets generated and sustained — through precession in binary systems, orbital motion, or plasma instabilities within the flow? What do their winding 
patterns reveal about jet stability and magnetic collimation on kiloparsec scales? If, instead, the RMs are dominated by the local ICM, corkscrew morphologies become powerful probes of cluster magnetism: what is the coherence scale and fluctuation spectrum of the ICM magnetic field, how 
does its strength vary with electron density, and does it align with jet-driven outflows or larger-scale cluster structure? More broadly, if quasi-periodic RM signatures can be detected and interpreted in the Corkscrew Galaxy, the same approach should be applicable to any sufficiently resolved radio galaxy whose jet samples a range of LOS depths through its local magnetised environment. As wide-field polarimetric surveys uncover growing numbers of extended tailed sources and radio galaxies with helical substructure, the methods developed here may provide a general framework for using radio galaxies as probes of ICM magnetic fields across a wide range of cluster environments. By disentangling these possibilities, we aim to determine whether corkscrew radio galaxies are valuable laboratories for studying both AGN jet physics and the magnetised ICM.

The paper is structured as follows. Section \ref{sec:obscal} describes details of the observations and calibration, Section \ref{sec:polim} details the polarimetric imaging techniques performed, Section \ref{sec:resultsandanalysis} outlines the analysis methods and results, and Section \ref{sec:discussion} and \ref{sec:conclusions} give the discussion and conclusions, respectively. More detailed methods can be found in Appendix \ref{sec:methodsappendix}. 

\begin{table*}
\caption{Summary of observations.}
\begin{threeparttable}
\centering
\begin{tabular}{ll}
\hline
\hline
No. of ASKAP pointings & 1\\
SBID (date) of observation & 11816 (2020-02-14) \\
Integration time & 8 hours \\
Full-band sensitivity in Stokes $Q$, $U$ \tnote{$\dagger$} & 18 $\muup$Jy beam$^{-1}$ / Stokes \\
Shortest baseline & 22.4 m \\
Longest baseline & 6.4 km \\
Angular resolution (robust $=0.0$) & 8$\times$8 \arcsec (smoothed) \\
Typical Largest recoverable angular scale\tnote{$\ddagger\ddagger$} & 30 arcmin\\
Frequency range  & 1296–1440 MHz \\
$\lambda^2$ range & 0.043--0.054 m $^2$\\
RM Spread Function FWHM \tnote{a} & 340 \radms \\
Largest recoverable RM-scale \tnote{$\ddagger\ddagger$} & 28 \radms \\
Largest recoverable |RM| \tnote{a} & 45,000 \radms \\
\hline
\end{tabular}

\begin{tablenotes}
\small
\item[$\dagger$] Measured per Stokes parameter in multi-frequency synthesis images generated with a Briggs' robust weighting value of 0.0.
\item[$\ddagger$] At the centre frequency of the band.
\item[$\ddagger\ddagger$] At greater than 50\% sensitivity, following \citet{Rudnick2023}.
\item[a] Calculated using RM-Tools version 1.3.1 \url{https://github.com/CIRADA-Tools/RM-Tools} \citep{Purcell2020,VanEck2026}, quoted to two significant figures.
\end{tablenotes}
\end{threeparttable}
\label{table:observational_details}
\end{table*}

\begin{table*}
\caption{Summary of properties of the Corkscrew Radio Galaxy, with columns (1) Source; (2) Sky position, given in right ascension and declination (J2000); (3) Galactic coordinates ($l, b$) in degrees to three decimal places; (4) Source size, expressed in both angular extent (arcminutes) and physical size (kiloparsecs); (5) Stokes 
\emph{I} intensities (25th, 50th, and 75th percentiles) in mJy/beam; (6) Host galaxy identification, based on the NASA/IPAC Extragalactic Database (NED); and (7) Redshift ($z$) values for the host galaxy, to four significant figures. Our adopted redshift was obtained from \citet{Woudt2008}.
}
\centering
\renewcommand{\arraystretch}{1.2}
\begin{tabular}{lccccccc}
\hline
 Source & RA, Dec & Galactic  & Size &  Stokes \emph{I} Pctl (25th, 50th, 75th) & Host Galaxy & Redshift  \\
 & (J2000) & $l, b$ (deg)  & (arcmin, kpc) &  (mJy/beam) & & \\
\hline
Corkscrew Galaxy & 16:14:57, $-$60:38:30   & 325.465, $-$7.014   & 28, 564   & 0.072, 0.16, 0.57   & ESO 137--G007   & 0.0165  \\
\hline
\end{tabular}
\label{table:individual_source_details}
\end{table*}

\section{Observations and Calibration}
\label{sec:obscal}
% \rb{ASKAP early science data supporting EMU, Wallaby and POSSUM survey definitions}
Our observations, summarised in Table~\ref{table:observational_details}, comprise ASKAP Band~2 (1296--1440~MHz) data obtained on 14~February~2020 during early-science observations for the POSSUM \citep{Gaensler2025}, EMU \citep{Norris2011}, and Wallaby \citep{Koribalski2020} surveys using ASKAP (\citealt{Johnston2007, DeBoer2009, Schinckel2016, Hotan2021}). A single 8~hour pointing was observed with the \texttt{square\_6x6} beam footprint \citep{McConnell2016}, employing a beam pitch of $0.9^{\circ}$.

Flux-scale calibration, bandpass calibration, and on-axis polarisation leakage (``D-term'') calibration were performed using the standard \textsc{askapsoft} pipeline\footnote{\textsc{askapsoft} v1.0.16 (\url{https://www.atnf.csiro.au/computing/software/askapsoft/sdp/docs/current/general/releaseNotes.html\#id14}).}, based on observations of the unpolarised calibrator PKS~B1934$-$638. The frequency-dependent instrumental XY-phase was corrected during beamforming using the on-dish calibration system \citep{Chippendale2019}.

\begin{figure*}
	\includegraphics[width=\textwidth]{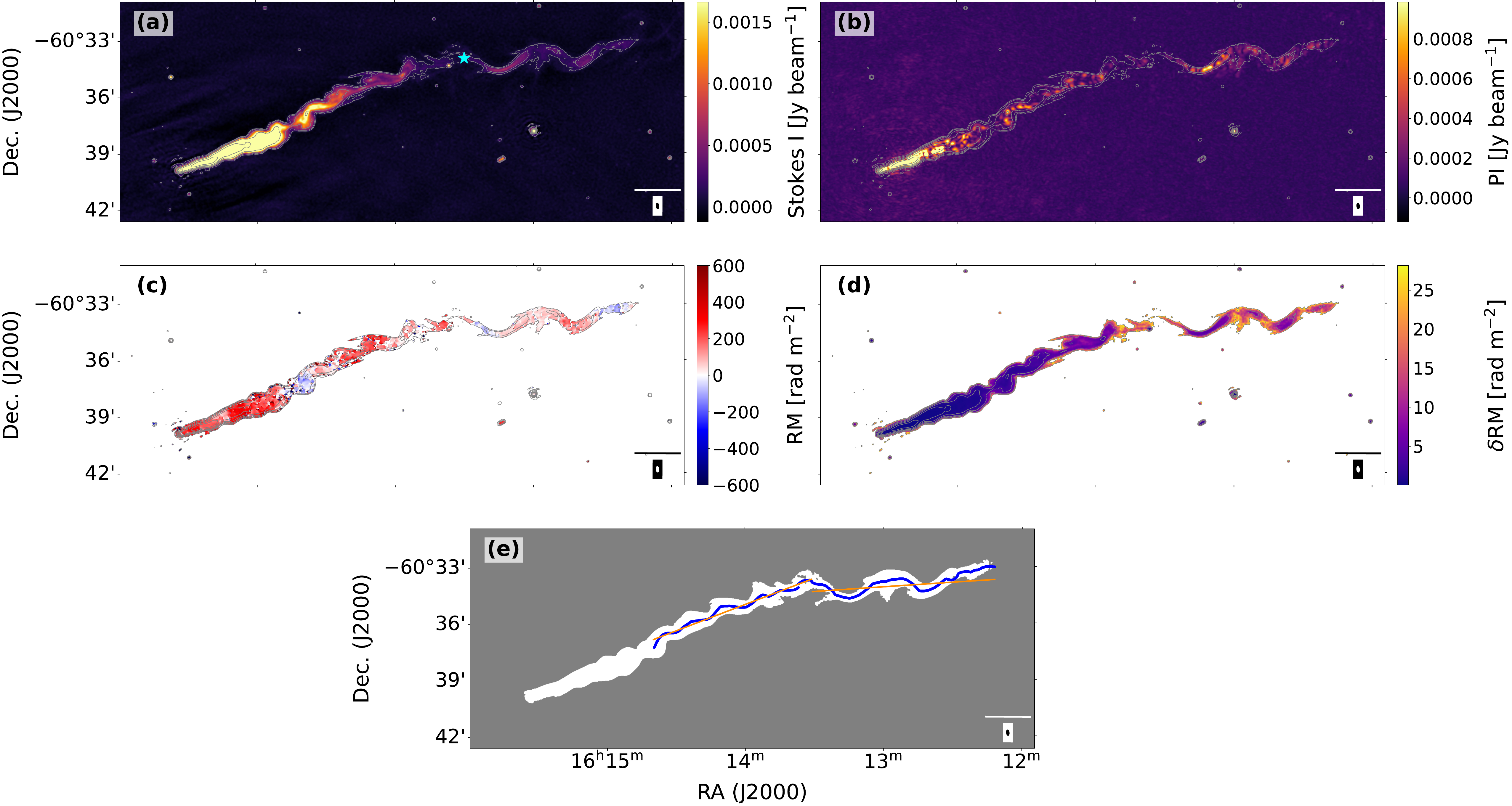}
    \caption{Descriptive maps of the Corkscrew Galaxy. \textbf{(a)} Stokes \emph{I} map with total intensity contours logarithmically spaced between $10^{-4}$ and $1$ Jy beam$^{-1}$. The cyan star marks the location where the jet bends and changes orientation (RA, Dec (J2000) $\approx 16^{\mathrm{h}}13.5^{\mathrm{m}}$, $-60^{\circ}34^{\prime}$). In our subsequent analysis, we consider the jet either side of this point separately. \textbf{(b)} Peak \emph{PI} map with total intensity contours as in panel (a). \textbf{(c)} Peak, foreground-corrected \emph{RM} map with total intensity contours as in panel (a). \textbf{(d)} Uncertainty in the peak RM with total intensity contours as in panel (a). \textbf{(e)} Source mask used for FilFinder analysis (white), the jet spine extracted using FilFinder (blue), and the local jet axis lines (orange). All panels share a common WCS coordinate system and as such, the x-axis in panel e) applies for all other panels. The beam size and a 100 kpc scale bar are displayed in the lower right on all panels.}
    \label{fig:descriptive_maps_corkscrew}
\end{figure*}

No correction was applied for the off-axis polarimetric response, as these observations predate the implementation of holography-based leakage modelling and subtraction \citep{Thomson2025,Gaensler2025}. Based on subsequent holography measurements and the fact that the regions of interest in our polarisation maps lie within 15\arcmin\ of the beam centre—well inside the $0.62^{\circ}$ half-width at half-maximum of the ASKAP primary beam at this frequency—we estimate the residual off-axis leakage from Stokes~$I$ into $Q$ and $U$ at the beam-centric radius of the Corkscrew Galaxy to be $\lesssim5$\,per~cent. For our adopted $7\sigma$ detection threshold ($\sigma_P = 18~\mu\mathrm{Jy\,beam^{-1}}$; see Section \ref{sec:polim}), detectable leakage would therefore arise only from Stokes~$I$ emission brighter than $\sim2.5~\mathrm{mJy\,beam^{-1}}$. Such emission is confined to the easternmost parts of the jet (Figure~\ref{fig:descriptive_maps_corkscrew}, panel~\emph{a}), where the corkscrew-like structure becomes unresolved and lies outside the region analysed in this work. Moreover, while residual leakage at this level may introduce small biases in the measured fractional polarisation, it cannot produce the frequency-dependent behaviour required to generate the RM structure observed toward the source (see Section~\ref{sec:jetstructure}).

\section{Polarimetric Imaging}
\label{sec:polim}
We imaged the calibrated measurement set using \textsc{WSClean} \citep{Offringa2014}. Imaging was performed using Briggs weighting with a robust parameter of 0.0, excluding baselines shorter than 370 wavelengths (\texttt{-minuv-l 370}) to eliminate residual radio and Solar interference on the shortest ASKAP baselines. An inner Tukey taper was applied to smooth this cut (\texttt{-taper-inner-tukey 800}). Images were produced for the closest beam, which was imaged individually from the larger $30$ $\mathrm{deg}^2$ field of view observed in SBID 11816, on a $3500 \times 3500$ pixel grid with a pixel scale of $1.5''$ and a padding factor of 1.5.

Phase-only self-calibration was performed in \textsc{CASA} \citep{CASATeam2022} using a Stokes~$I$ continuum model for two rounds with a solution interval of 300 seconds, followed by a single round of combined phase-and-amplitude self-calibration with a solution interval of 600 seconds. Shorter solution intervals were tested but did not yield significant improvements.

Following self-calibration, final frequency-dependent polarimetric imaging was performed using \textsc{WSClean}. Deconvolution was carried out using \texttt{-join-polarizations} together with \texttt{-link-polarizations qu}, linking the Stokes $Q$ and $U$ polarisation products during cleaning. Frequency channels were imaged using \texttt{-squared-channel-joining}, producing frequency-resolved Stokes $Q$ and $U$ image cubes. A Stokes $I$ data cube was separately generated in an analogous fashion.

All frequency channels were convolved to the spatial resolution of the lowest-frequency channel ($8 \times 7.5$~arcsec at position angle $78^{\circ}$), regridded to a common pixel grid, and concatenated to form Stokes $I$, $Q$, and $U$ data cubes as functions of right ascension, declination, and $\lambda^2$. For total-intensity analysis and visualisation, a higher-sensitivity Stokes $I$ multi-frequency synthesis (MFS) image was generated independently.

RM synthesis was applied to the $IQU$ data cubes using \textsc{RM-Tools}\footnote{\textsc{RM-Tools} v1.3.1; \url{https://github.com/CIRADA-Tools/RM-Tools}} \citep{VanEck2026}. From the peak of the Faraday dispersion function (FDF; \citealt{Brentjens_deBruyn_2005}), maps of RM and polarised intensity (\emph{PI}) were extracted. RM uncertainties were estimated as $\delta \mathrm{RM} = W/(2\,\mathrm{SNR})$, where $W$ is the full width at half maximum of the RM spread function (Table~\ref{table:observational_details}) and SNR is the polarised signal-to-noise ratio. A polarised SNR threshold of $\mathrm{SNR} \geq 7$ was applied to mask the \emph{PI} and RM maps for subsequent analysis.

We apply a first-order Galactic foreground correction following \citet{Anderson2018}, subtracting the median RM measured across the masked source region from all RMs ($\mathrm{RM}_{\mathrm{fg}} \approx -19$ \radms). We note that this value is broadly consistent with the Galactic RM foreground model of \citet{Hutschenreuter2022}, which predicts $\mathrm{RM}_{\rm fg} = (32 \pm 44)$ \radms ~at this location, noting that their model primarily probes foreground structure on degree scales. Unless otherwise stated, all RM values and maps presented in this paper include this correction. Residual foreground contributions are assessed in Section \ref{sec:GalFG}.

\section{Results}
\label{sec:resultsandanalysis}

\subsection{Jet Morphology and Polarimetric Structure}
\label{sec:jetstructure}

The Stokes \emph{I}, \emph{PI},  \emph{RM}, and $\delta RM$ maps are shown in Figure~\ref{fig:descriptive_maps_corkscrew} panels~(a), (b), (c), and (d), respectively. A detailed description of the Stokes~$I$ morphology of this source is presented by \citet{Koribalski2024}; here we summarise the key features relevant to the present analysis. The total intensity emission is brightest along the central jet spine and fades smoothly towards the edges. The brightest emission is located on the eastern region of the jet, coincident with the central AGN and host galaxy, reaching a peak of 64.7~mJy\,beam$^{-1}$. Toward the western end of the jet, the emission declines to approximately 0.4~mJy\,beam$^{-1}$.

The \emph{PI} map exhibits a combination of compact bright knots embedded within more diffuse, lower-surface-brightness emission that broadly traces the Stokes~$I$ morphology. Enhanced \emph{PI} is observed toward the core on the eastern region of the jet, as well as in the fainter western region immediately downstream of the apparent jet disruption and bending (RA, Dec (J2000) $\approx 16^{\mathrm{h}}13.5^{\mathrm{m}}$, $-60^{\circ}34^{\prime}$). 
% Across the source, \emph{PI} values range from approximately 0.03~mJy\,beam$^{-1}$ to 10.8~mJy\,beam$^{-1}$.

The RM map reveals alternating regions of positive and negative values, spanning hundreds of $\radms$. These RM variations persist over angular scales comparable to the characteristic windings of the jet seen in Stokes~$I$. This correspondence motivates an assessment of whether the observed RM structure could arise from foreground Faraday rotation rather than being intrinsic to the source. In the following section, we therefore examine the contribution of the Galactic interstellar medium to the observed RM variations.

\begin{figure}
	\includegraphics[width=\columnwidth]{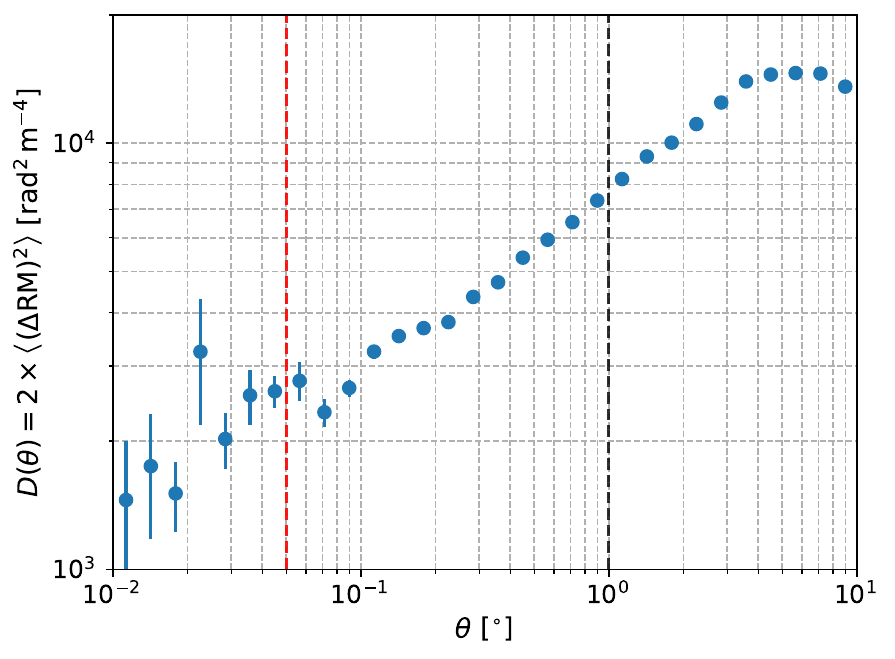}
    \caption{
RM structure function of the field surrounding the Corkscrew Galaxy, computed from the POSSUM RM grid within $\pm1^{\circ}$ in Galactic latitude and $\pm10^{\circ}$ in Galactic longitude of the source. Bins are logarithmically spaced, spanning $0.01$--$10^{\circ}$ in angular separation. Noise contribution arising from measurement uncertainties has been subtracted from each bin. Vertical error bars show the standard error of the mean within each bin. The dashed black line marks the one-degree scale associated with Galactic RM fluctuations, while the red dashed line indicates the angular scale corresponding to the typical spacing of the Corkscrew jet windings (a few arcminutes).}
    \label{fig:structurefunc}
\end{figure}

\begin{figure}
	\includegraphics[width=\columnwidth]{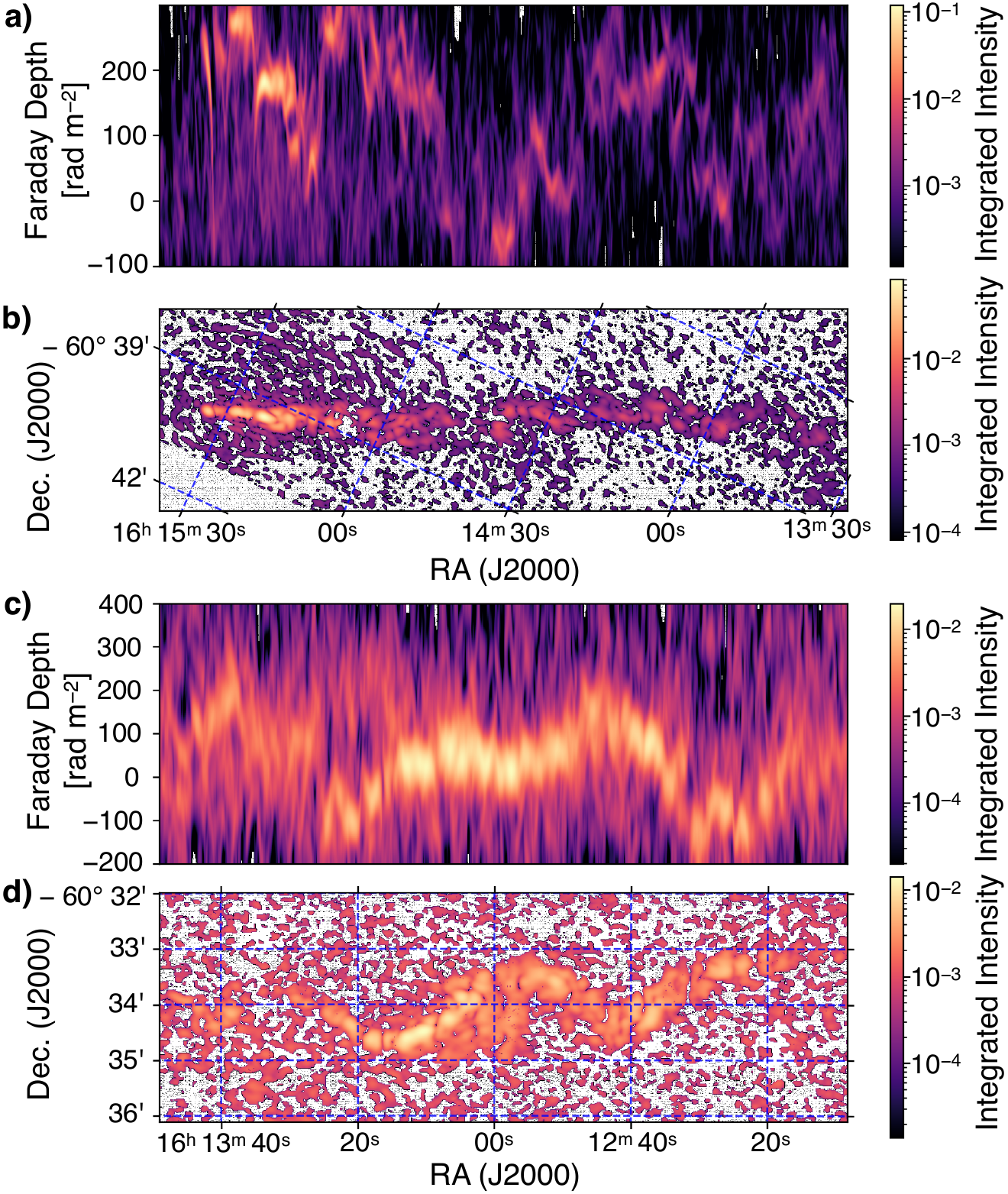}
    \caption{2D projections of the pseudo-3D polarisation cube for the Corkscrew Galaxy generated
    using the technique of \citet{Rudnick2024}, introduced in Section \ref{sec:results_pseudo3D}. a) shows the projection of the pseudo-3D polarisation cube onto the (RA, \emph{RM}) plane for the east-most part of the Corkscrew Galaxy jet, while b) shows the \emph{PI} for this region. c) and d) show the same projections as a) and b), but for the west-most part of the Corkscrew Galaxy jet. In a), the RM range goes from $-$100 to $+$290 \radms, while in c) it goes from $-$200 to $+$400 \radms. The smoothing width is 3 \radms. To generate b) and d), the pseudo-3D polarisation cube was rotated by a specified angle in the plane-of-sky about the source's geometric centre to allow for comparison of spatial features with the RM projections.}
    \label{fig:corkscrew_2dprojections}
\end{figure}

\subsection{Residual Galactic Rotation Measure Foreground}
\label{sec:GalFG}
Large-amplitude RM fluctuations ($\sim100~\radms$) are known to occur on degree scales near the Galactic plane, but both their amplitude and coherence length decrease rapidly with increasing Galactic latitude and decreasing angular scale (e.g.~\citealt{Schnitzeler2010,Oppermann2012,Hutschenreuter2022,Ma2025}). Although a constant Galactic foreground correction of $-19~\radms$ has already been applied (Section~\ref{sec:polim}), we explicitly test whether the smaller-scale RM variations described in Section~\ref{sec:jetstructure} could plausibly arise from residual Galactic foreground structure at the location of the Corkscrew Galaxy ($|b|\simeq7^{\circ}$).

We compute the second-order RM structure function of the surrounding field using unpublished RM grid data from the POSSUM survey \citep{Vanderwoude2024,Gaensler2025}, restricting the sample to sources within $\pm1^{\circ}$ in Galactic latitude and $\pm10^{\circ}$ in Galactic longitude (see Appendix~\ref{sec:rm_reliability} for methodological details).
The resulting structure function (Figure~\ref{fig:structurefunc}) shows RMS RM fluctuations of order $\sim100~\mathrm{rad\,m^{-2}}$ on degree scales, decreasing to $\sim30~\mathrm{rad\,m^{-2}}$ on the  angular scales characteristic of the jet windings (corresponding to a $\sim$few arcminutes; see Figure~\ref{fig:descriptive_maps_corkscrew}, panel~a; Section~\ref{sec:periodicity}). At these scales, the foreground RM fluctuations are more than an order of magnitude smaller than the RM variations observed across the Corkscrew Galaxy (Figure~\ref{fig:descriptive_maps_corkscrew}, panel~c). We therefore conclude that residual Galactic foreground Faraday rotation cannot account for the RM structure observed toward the source.

\subsection{RM--Morphology Correspondence via Pseudo-3D Visualisation}
\label{sec:results_pseudo3D}

While patchy positive and negative RM structure was noted in Section~\ref{sec:jetstructure}, it is difficult to assess the spatial relationship between Faraday depth and the projected jet morphology by comparing the plane-of-sky maps alone. \citet{Rudnick2024} introduced the pseudo-3D visualisation technique to address this limitation by explicitly propagating the position--position--\emph{RM} information into a cube representation, which can then be re-projected along any choice of axis to reveal structure visually.

Following this approach, we produce two-dimensional projections in both plane-of-sky coordinates (RA, Dec.) and Faraday depth space (RA,\emph{RM}), shown in Figure~\ref{fig:corkscrew_2dprojections}. Panels~(a) and (b) display the RM and \emph{PI} projections, respectively, for the eastern portion of the jet, while panels~(c) and (d) show the corresponding projections for the western portion. The RM ranges span $-100$ to $+290~\radms$ in panel~(a) and $-200$ to $+400~\radms$ in panel~(c), with a smoothing width of 3~\radms. For the \emph{PI} projections in panels~(b) and (d), the pseudo-3D cube was rotated by a fixed angle in the plane-of-sky about the geometric centre of the source to facilitate comparison between projected jet structure and Faraday depth variations.

In these projections, the RM distribution along the jet exhibits an apparently regular oscillatory pattern that visually resembles the plane-of-sky oscillation traced by the jet morphology in \emph{PI}. This correspondence is most evident in the western portion of the jet, where successive RM extrema occur at positions along the jet axis similar to the spacing of the morphological windings, but a comparable pattern is also apparent in the eastern portion of the jet. The pseudo-3D representation therefore suggests a systematic correspondence between RM variations and the corkscrew morphology of the jet. In the following section, we examine this apparent correspondence quantitatively by measuring the periodicity of the RM signal and its phase relationship with the jet deviation using one-dimensional statistical analyses.

\subsection{Quantitative analysis of RM–jet correspondence}\label{sec:results_filfinder}

\begin{figure*}
	\includegraphics[width=\textwidth]{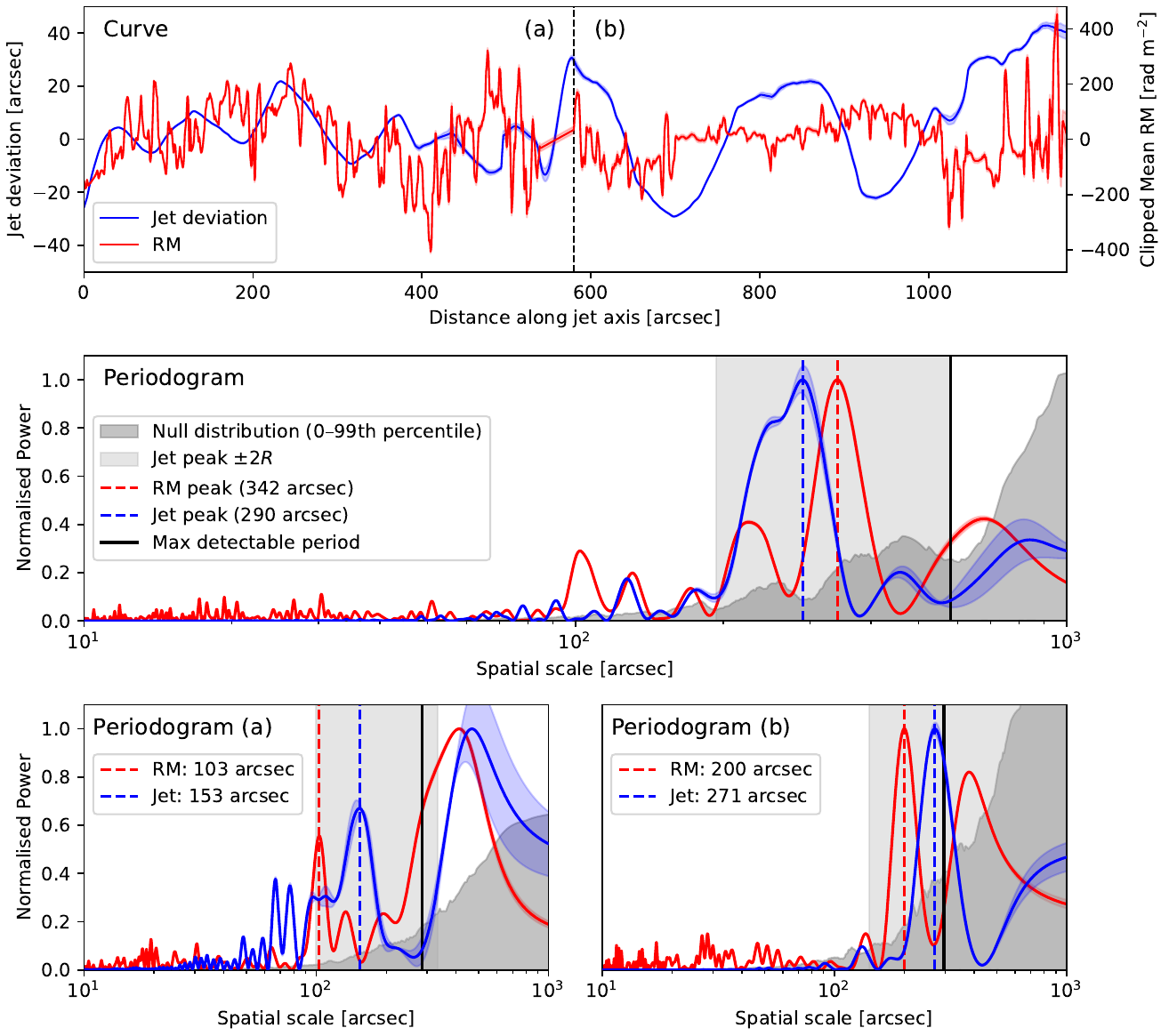}
    \caption{Periodogram analysis for jet deviation and RM. \textbf{Top:} Jet deviation (blue) and RM (red) as a function of distance along the jet axis, with shaded 1$\sigma$ uncertainties. The black dashed vertical line marks the separation between the tightly-wound (a) and loosely-wound (b) portions of the jet. \textbf{Middle:} Lomb--Scargle periodogram of the full jet for both quantities, normalised to their respective peak powers. Monte Carlo realisations of both the jet deviation and RM profiles quantify the $1\sigma$ variability expected from measurement uncertainties (blue and red shaded regions). The dark grey shaded region denotes the 0-99th percentile range of the 1000 null RM realisations normalised using the same normalisation factor as for the observed RM periodogram (see Section \ref{sec:periodicity}). Dashed lines indicate the strongest detected jet deviation and RM periods, with the Rayleigh limit $\pm 2R$ window (centred on the jet deviation peak) shown in light grey. The maximum detectable period is denoted with a solid black vertical line. \textbf{Bottom:} Periodograms computed separately for sections (a) and (b), each plotted with corresponding Monte-Carlo error bands and 0-99th percentile range of null RM realisations. The peak power for each quantity is indicated with a dashed vertical line. While omitted from the legend, the Rayleigh limit $\pm 2R$ window and maximum detectable period are displayed as for the middle panel.}
    \label{fig:curves_periodograms}
\end{figure*}

\subsubsection{Measuring Jet Deviation and RM Along the Jet}\label{filfindermethodandresults}

To enable quantitative comparison between the corkscrew jet morphology and the associated RM structure, we first construct one-dimensional representations of both quantities along the jet axis. This step transforms the projected jet trajectory and the RM distribution into spatial series defined on a common coordinate, providing the basis for subsequent periodicity and correlation analyses. We use the \textsc{FilFinder} package \citep{Koch2015} to trace the central spine of the jet from the Stokes~$I$ continuum image. Panel~(e) of Figure~\ref{fig:descriptive_maps_corkscrew} shows the binary source mask used for the \textsc{FilFinder} analysis together with the resulting jet spine (blue). The extracted spine was smoothed using a locally weighted regression (LOWESS) to produce a continuous representation of the jet trajectory. Hereafter, when we refer to the `eastern' portion of the jet, we explicitly refer to the `eastern-most' portion of our analysis region rather than the eastern-most part of the jet. 

Jet deviation was quantified by measuring lateral displacements perpendicular to a representative jet axis, defined separately for the eastern and western segments to account for the clear change in jet orientation near RA (J2000) $\sim16^{\mathrm{h}}13.5^{\mathrm{m}}$. These representative axes are shown in orange in panel~(e) of Figure~\ref{fig:descriptive_maps_corkscrew}. Monte Carlo perturbations of the axis position were used to estimate systematic uncertainties in the resulting jet deviation curves (further details are provided in Appendix \ref{sec:methodsappendix}).

To construct a corresponding RM profile, trimmed-mean RM values were extracted from narrow rectangular strips oriented perpendicular to the local jet direction and sampled at regular intervals along the jet spine. Uncertainties were estimated via Monte Carlo realisations incorporating per-pixel RM errors. The jet deviation and RM profiles were interpolated onto a common spatial grid along the jet axis and detrended to isolate oscillatory structure for subsequent correlation and spectral analyses.

The resulting jet deviation and RM profiles are shown in the upper panel of Figure~\ref{fig:curves_periodograms}. Both exhibit quasi-periodic structure on comparable spatial scales, motivating a quantitative assessment of their characteristic periods and relative phase, which we present below.

\subsubsection{Periodicity analysis}
\label{sec:periodicity}
To test whether the jet deviation and RM series contain statistically significant spatial periodicities, we compute Lomb--Scargle periodograms \citep{Lomb1976,Scargle1982,VanderPlas2018} for each curve. This analysis determines whether either signal exhibits dominant oscillatory scales and whether these scales are consistent between the jet morphology and the RM structure, both for the full jet and for the two spatial regions separated by the morphological transition described in Section \ref{sec:intro} and shown in Figure~\ref{fig:curves_periodograms}.

The middle panel of Figure~\ref{fig:curves_periodograms} presents the Lomb--Scargle periodograms for the full jet. Both the jet deviation and RM periodograms display strong peaks, with the latter well above those obtained from an ensemble of 1000 simulated foreground RM profiles (dark grey shaded region). These null profiles are constructed by generating Gaussian random field (GRF) realisations of Galactic foreground RMs, with spatial correlation properties derived from the observed RM structure function of background sources within $\pm1^{\circ}$ in Galactic latitude and $\pm10^{\circ}$ in Galactic longitude of the Corkscrew Galaxy. The structure function is well described by a power law $D(\theta) \propto \theta^{\alpha}$ with $\alpha = 0.48$, saturating at a correlation length of $ \sim 200$ arcmin. Each GRF realisation is generated on a grid matching the spatial extent and resolution of the observed RM map. For each realisation, a trimmed-mean RM profile is extracted in the same manner as was done for the observed RM curve. The Lomb--Scargle periodogram of each simulated profile is then computed and normalised to the peak power of the observed RM periodogram, forming the null distribution against which the observed RM peaks are assessed. 

To ensure reliable interpretation of the periodograms, we follow the recommendations of \citet{RamirezDelgado2025} and apply the Rayleigh criterion to define the resolution limits imposed by the finite length of the jet (1163 arcseconds for the full jet, 570 arcseconds and 593 arcseconds for the eastern and western portions of the jet, respectively). For the full jet, the Rayleigh frequency is $R = 1/T = 0.00086~\mathrm{arcsec^{-1}}$, corresponding to a maximum detectable period of $P_{\max} = 1/(2R) = 581.6$~arcsec. The Rayleigh frequency also sets the effective frequency resolution, which we propagate into uncertainties on the measured periods.

Within these limits, the jet deviation and RM periodograms show dominant peaks at comparable spatial scales, with characteristic periods of $(290 \pm 72)$~arcsec for the jet deviation and $(342 \pm 101)$~arcsec for the RM signal, the amplitude of the latter being 11 times higher than the 84th percentile of the null distribution. The shaded light grey bands in Figure~\ref{fig:curves_periodograms} highlight the $\pm2R$ interval around the dominant jet-deviation period, within which the RM peak also lies.

The bottom panels of Figure~\ref{fig:curves_periodograms} show periodograms computed separately for regions~(a) (eastern portion of the jet) and~(b) (western portion). In region~(a), both profiles exhibit shorter-period oscillations, with dominant peaks at $(153 \pm 41)$~arcsec for the jet deviation and $(103 \pm 19)$~arcsec for the RM signal, the latter 60 times the amplitude of the 84th percentile of the null ensemble. In region~(b), both curves show stronger power at longer spatial scales, with dominant peaks at $(271 \pm 124)$~arcsec for the jet deviation and $(200 \pm 67)$~arcsec for the RM, rising to 13 times the amplitude of the 84th percentile of the null ensemble.

We therefore find that the jet deviation and RM profiles share quantitatively similar spatial periodicities within the resolution limits of the data, and that these periodicities cannot be reproduced by Galactic RM foreground fluctuations alone. Having established this correspondence in the frequency domain, we now examine whether the two signals are also coupled in phase along the jet using cross-correlation analysis.

\begin{figure*}
	\includegraphics[width=\textwidth]{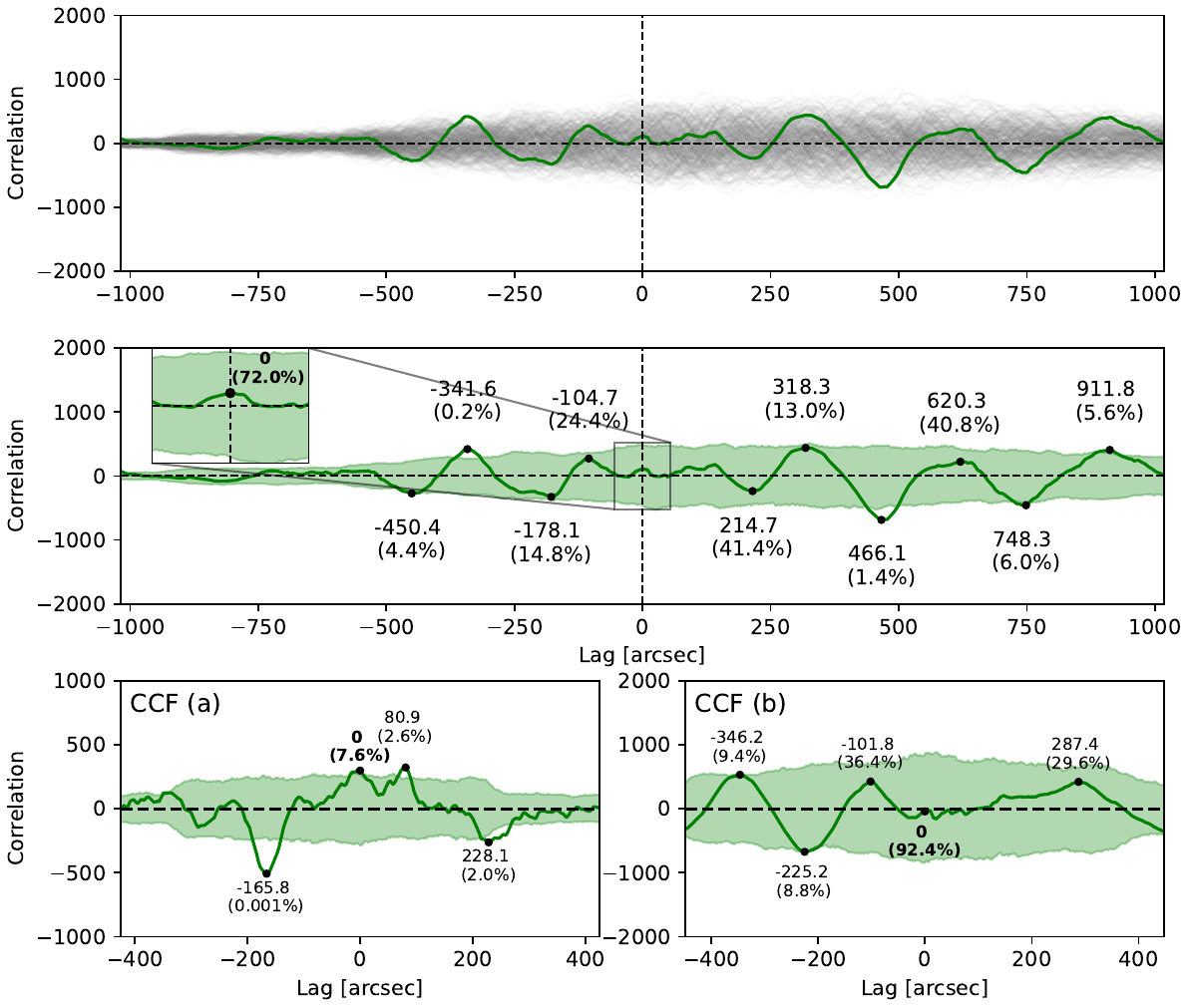}
    \caption{Cross-correlation analysis between the jet deviation and RM curves. \textbf{Top:} The CCF of the jet deviation and RM curves (green) compared with null realisations (grey) that show the range of correlations expected by chance between the jet deviation curve and phase-randomised RM curves. Only lags within $\pm$ 1000 arcsec are shown to ensure that the CCF limits the number of overlapping points to no less than half of the dominant spatial scale determined from the periodogram analysis. Vertical and horizontal dashed lines represent zero lag and zero correlation, respectively.
    \textbf{Middle:} The CCF with the 5th--95th percentile interval of the null distribution (the range containing 90\% of null CCFs; green shaded region). Peaks identified using a prominence-based peak finder are marked with the lag at which they occur, and the percentage of nulls that exceed the true CCF value (in parentheses). The inset zooms into the lag range of $\pm 55$ arcsec, highlighting the percentage of nulls that exceed the true CCF value at zero lag.
    \textbf{Bottom:} CCFs computed separately for sections (a) and (b) of the jet (defined in Figure \ref{fig:curves_periodograms}), each compared with their respective 5th--95th percentile null envelopes. Both bottom panels have prominent peaks annotated as for the middle panel, as well as the percentage of nulls that exceed the true CCF values at zero lag (bolded).}
    \label{fig:cross_correlation}
\end{figure*}

\subsubsection{Spatial Correlation Analysis}\label{sec:correlations}
Having established that the jet deviation and RM series exhibit comparable spatial 
periodicities, we next examine how these oscillations are phased relative to one 
another using the cross-correlation function (CCF). The CCF quantifies the 
correspondence between the two curves as a function of spatial lag. The zero-lag correlation
value is of primary interest: a high correlation at zero lag would suggest that RM and jet deviation extrema coincide spatially, whilst a value consistent with zero 
would suggest a phase offset between the two signals. Full details of the CCF 
computation, null test construction, peak-finding procedure, and lag range 
restrictions are given in Appendix~\ref{sec:ccf}.

We first note an important expectation: if either signal alone contains a periodic component, the CCF will exhibit a regular sequence of 
alternating peaks and troughs at multiples of the characteristic spatial scale, 
regardless of whether the two signals are physically related. The presence of 
such a pattern is therefore not sufficient evidence of a genuine cross-correlation. Here, the zero-lag value is the primary diagnostic of alignment, whilst the 
regularity of the non-zero-lag features serves to check for consistency with the shared periodicity identified in Section~\ref{sec:periodicity}.

To assess whether the observed zero-lag correlation is physically meaningful, we 
construct a null distribution from 500 synthetic RM curves generated with the 
Timmer--K\"{o}nig method \citep{TimmerKoenig1995}. These null RM curves preserve the 
power spectrum of the observed RM signal — and hence its autocorrelation structure 
and amplitude distribution — while randomising its phase structure. Each null curve is 
cross-correlated with the observed jet deviation curve to produce a null CCF. The null distribution therefore represents the full ensemble of CCF amplitudes 
achievable given the power spectra of both signals, across many possible phase arrangements of the RM. We note that by construction, it is impossible for the true CCF peaks to exceed the null distribution envelope, because the observed CCF is a realisation drawn from the same ensemble. Additionally, because the number of 
independent phases in a finite series is limited, it is possible that some surrogate phase arrangements will produce CCF amplitudes exceeding the true curve at particular lags by chance. For these reasons, the null test should therefore not be interpreted as a 
threshold to be exceeded, but rather as a means of establishing the range of CCF 
amplitudes consistent with the observed power spectra alone. If the true CCF ranks 
amongst the highest of the ensemble at a given lag, this indicates that the observed 
phase arrangement of the RM is unusually well phase-aligned with the jet windings — a 
result that would be unlikely to occur if the two signals shared no coherent phase relationship. The converse (i.e. the true zero-lag CCF amplitude ranking close to zero compared to the ensemble at a given lag) likewise suggests the RM and jet deviation curves are not phase-synced. Prominent CCF features are identified using a prominence threshold corresponding to the $\sim$95th percentile of the null prominence distribution, ensuring that only 
clear peaks are retained.
% If the true CCF ranks 
% amongst the highest of the ensemble at zero lag, this would suggest that the observed 
% phase arrangement of the RM is unusually well-aligned with the jet periodicity.

The CCF results are shown in Figure~\ref{fig:cross_correlation}. When applied over the full extent of the jet
(middle panel), the zero-lag CCF is exceeded by $72.0\%$ of the null ensemble. This suggests that when the jet is considered as a whole, the RM and jet 
deviation are weakly consistent with a phase-shifted relationship rather than direct spatial 
alignment. However, splitting the jet at the morphological transition defined in 
Figure~\ref{fig:curves_periodograms} reveals distinct behaviour in each region (bottom panels a) and b)), with an observed difference in the 
zero-lag correlation between the two sections. In section~(a), the zero-lag CCF is 
exceeded by just $7.6\%$ of null curves, suggesting that the RM and jet deviation extreme are relatively phase-aligned 
in this region. In jet section~(b), by contrast, the zero-lag 
CCF is exceeded by $92.4\%$ of nulls, suggesting that the two curves are more likely to be
offset in phase. We caution that these 
sectional results are based on shorter data segments and should be interpreted tentatively, but nevertheless, the contrast between the two sections is suggestive of different phase relationships on either side of the transition in morphology of the jet windings and projected orientation of the jet axis.

The disparity between the sectional and full-jet results highlights the susceptibility of CCF analyses to cancellation effects. If different regions of the 
jet exhibit correlations with differing phase relationships, combining them suppresses 
the zero-lag signal even when individual regions show meaningful correspondence. 
Splitting at the morphological transition is therefore essential to avoid conflating independent behaviours. However, as mentioned, shorter segments reduce the available data points and restrict the interpretable lag range, limiting the robustness of non-zero-lag 
features in the sectional CCFs.

We therefore extend the analysis to the non-zero-lag features in the full-jet CCF. This curve exhibits a sequence of alternating peaks and 
troughs at characteristic lags (Figure~\ref{fig:cross_correlation}; 
Table~\ref{tab:ccf_extrema}). The mean separation between successive same-sign 
features is $273$~arcsec, with a standard deviation of $23$~arcsec — a scatter of $\sim$$8\%$ of the mean separation. This regularity is consistent with 
both signals sharing a common underlying spatial scale, and the mean separation is in agreement with the dominant spatial scale of $(290 \pm 72)$~arcsec identified 
in the periodogram analysis (Section~\ref{sec:periodicity}). We interpret this 
as supporting evidence for our claim of shared periodicity and a phase relationship between the jet deviation and RM curves.

Summarising the main result from the CCF analysis, the sectional zero-lag results tentatively suggest that the RM and jet deviation may be more closely aligned in the low-amplitude region~(a) than in 
the high-amplitude region~(b), where a phase offset appears more likely. 
% The 
% first prominent non-zero-lag CCF peak occurs at $\sim$$-105$~arcsec, corresponding 
% to a phase offset of roughly $105/290 \approx 0.4$ cycles, which may reflect the 
% typical phase relationship between the two oscillations in the regions where they are 
% not in phase.

\begin{table}
\centering
\caption{Prominent non-zero-lag extrema in the CCF between the jet deviation and RM 
curves for the full jet shown in Figure~\ref{fig:cross_correlation}. Listed extrema 
exceed the adopted prominence threshold and are characterised by their spatial lag 
and the fraction of null CCF realisations that exceed the observed amplitude (null 
exceedance fraction). Individual lag values are not interpreted independently; 
rather, the regularity of the spacing between successive same-sign features — with a 
standard deviation of $23$~arcsec about a mean of $273$~arcsec — is consistent with 
the characteristic spatial scale identified in the periodogram analysis.}
\label{tab:ccf_extrema}
\begin{tabular}{lcc}
\hline
Extremum type & Lag (arcsec) & Null exceedance (\%) \\
\hline
Peak   & $-341.6$ & $0.2$ \\
Peak   & $-104.7$ & $24.4$ \\
Peak   & $+318.3$ & $13.0$ \\
Peak   & $+620.3$ & $40.8$ \\
Peak   & $+911.8$ & $5.6$ \\
\hline
Trough & $-450.0$ & $4.4$ \\
Trough & $-178.1$ & $14.8$ \\
Trough & $+214.7$ & $41.4$ \\
Trough & $+466.1$ & $1.4$ \\
Trough & $+748.3$ & $6.0$ \\
\hline
\end{tabular}
\end{table}

\subsubsection{Results Summary}\label{sec:results_summary}

Summarising our main results, we find that the RM maps show alternating positive and negative values along the Corkscrew Galaxy jet, with amplitudes reaching approximately several hundred \radms and characteristic angular scales of several arcminutes, which is comparable to the observed jet windings. Using the POSSUM RM grid, we measure Galactic foreground RM fluctuations on these spatial scales that are an order of magnitude smaller than those observed across the jet and therefore do not reproduce the measured RM structure.

From one-dimensional profiles extracted along the jet spine, we measure statistically significant spatial periodicities in both the RM and jet-deviation signals. Our Lomb–Scargle periodogram analysis identifies dominant RM periods that are comparable, within uncertainties and resolution limits, to those measured for the jet deviation, for both the full jet and the eastern and western jet segments analysed separately. Using Galactic foreground-dominated RM realisations, we find that the observed RM periodogram peaks are not generated by the Galactic foreground alone, and exceed the null ensemble by at least 11 times the 84th percentile of the null distribution.

We further compute CCFs between the RM and jet-deviation profiles and identify a repeating sequence of extrema at spatial separations consistent with the dominant periods measured for the RM and jet-deviation in the periodogram analysis. The zero-lag correlation for the full jet is low in amplitude compared to these extrema, and low in amplitude compared to the envelope of null testing curves. We find that the same holds true for the western part of the jet considered in isolation, but that the CCF for the eastern part of the jet shows the opposite --- that is, comparatively high correlation at zero lag.

\section{Discussion}
\label{sec:discussion}

\subsection{Faraday Rotation In the Vicinity of Large-Scale Radio Jets}
\label{sec:faraday_rotating_zone_around_radio_galaxies}

We have identified oscillatory RM variations along the Corkscrew jet (Sections \ref{sec:jetstructure} and \ref{sec:results_pseudo3D}) with periods of $\sim$ several arcminutes (Section \ref{sec:periodicity}).
In Section~\ref{sec:GalFG}, we demonstrated that Galactic foreground fluctuations cannot account for the amplitude of these RM variations at this scale. We now therefore consider whether such structure could plausibly be produced by the physically unrelated ICM in the foreground of the jet.

In standard models of the ICM, magnetic field and electron density decline smoothly with radius, with superposed fluctuations that are typically described as an isotropic, Gaussian random field with a Kolmogorov-like power spectrum \citep{Donnert2018}. Under these conditions, foreground ICM Faraday rotation produces spatially correlated but non-repeating RM fluctuations, with no preferred phase coherence or tendency to track the morphology of individual radio sources. Such a foreground would not naturally generate RM oscillations that repeat at fixed spatial intervals or correlate systematically with the winding structure of an embedded jet.

Recent POSSUM studies have revealed that cluster outskirts can exhibit departures from simple azimuthally-symmetric models, including non-monotonic RM enhancements associated with merger activity \citep{Anderson2021, Osinga2022, Osinga2025, Loi2025, Khadir2025}, including strong RM gradients. However, reproducing the observed RM behaviour in the Corkscrew Galaxy via an ICM foreground would require an implausible combination of conditions where the ICM structure coincidentally aligns with the radio jet axis. We are not aware of any evidence that supports such a configuration in the Norma cluster.  

% a merger-driven magnetic structure aligned with the jet axis, RM sign reversals across that axis, and persistence of the RM oscillations across the observed jet orientation break. No evidence currently supports such a configuration in the Norma cluster.  

We therefore conclude that the observed quasi-periodic RM variations cannot be attributed to the unassociated ICM foreground alone, though such a contribution could explain the low-amplitude `jitter' superposed on the broader variations observed in Figures \ref{fig:corkscrew_2dprojections} (panels a \& c) and \ref{fig:curves_periodograms} (top panel). Instead, the spatial correspondence between RM structure and jet morphology points to a Faraday-rotating region that is physically associated with, or located local to the jet itself. In the following section, we examine the physical nature of this Faraday-rotating plasma and its implications for jet–environment interactions.

\subsection{Jet, Sheath, or Enclosed ICM?}
\label{sec:physical_nature_of_faraday_rotating_region}

The structured relationship between RM variations and jet deviations established in Sections~\ref{sec:periodicity} and~\ref{sec:correlations} provides direct constraints on the physical origin of the Faraday-rotating plasma in the Corkscrew Galaxy. Having ruled out both Galactic and large-scale ICM foregrounds as dominant contributors, we interpret the observed RM signal as arising from plasma local to the jet environment. Here we assess the physical nature of this Faraday-rotating region by comparing two limiting classes of models: (i) Faraday rotation generated by magnetised plasma internal to the jet or in a sheath immediately surrounding the jet, and (ii) Faraday rotation arising primarily in the local ICM immediately surrounding the jet, with the jet acting as a geometrical probe of path-length variations. These scenarios represent end members of a continuum, and the relative importance of each may vary along the jet length.

\begin{figure*}
	\includegraphics[width=\textwidth]{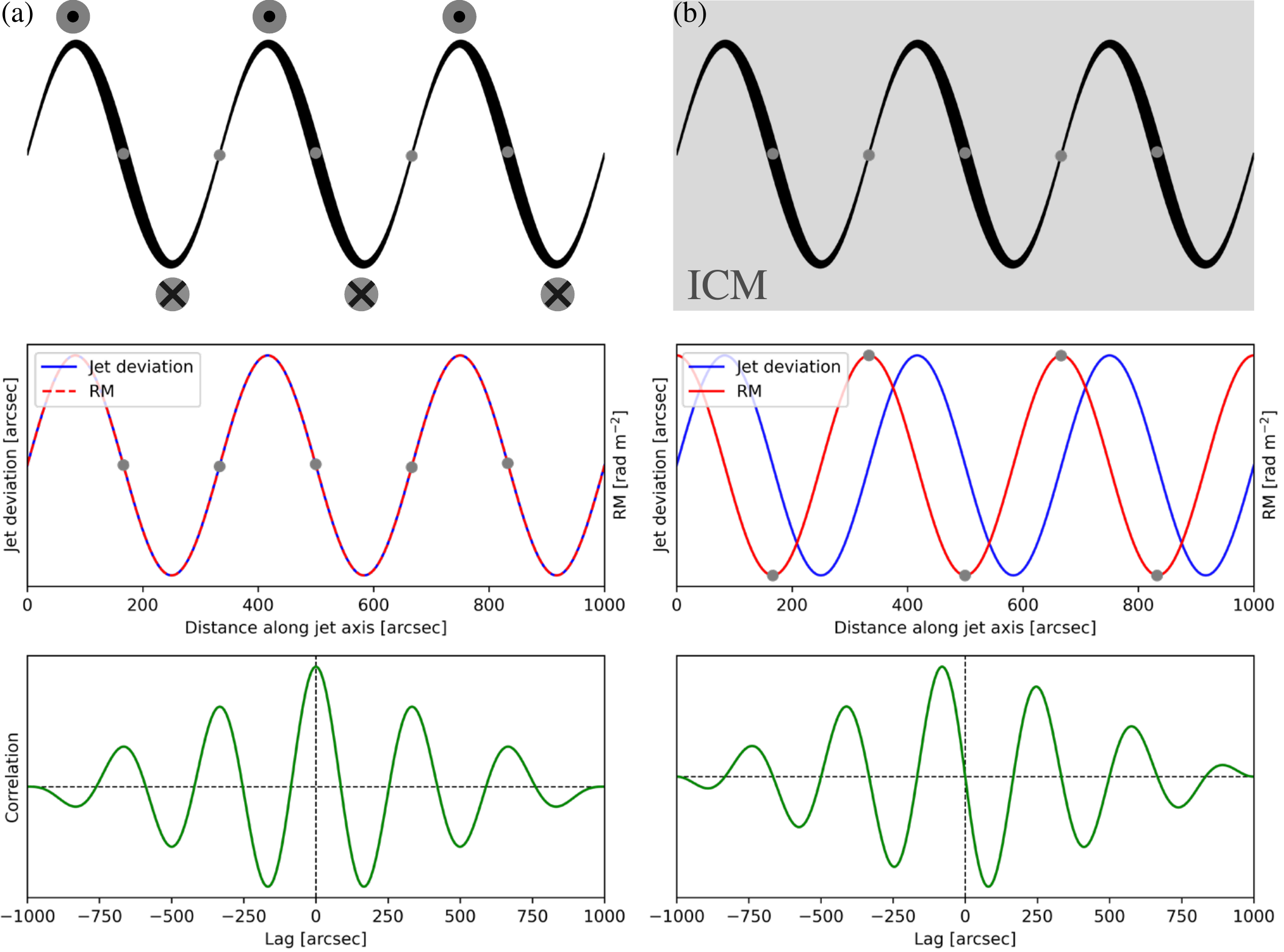}
    \caption{Schematic depicting a toy model for two possible Faraday-rotating regions: (a) the Faraday-rotating region is physically associated with the jet, or (b) the RMs are generated in the ICM in the near vicinity of the jet (e.g., the in-situ ICM within the jet windings). Each panel shows: (top) a schematic of the scenario as seen on the plane of the sky, with the observer looking in from out of the page; (middle) simplified jet deviation and RM curves corresponding to the scenario; and (bottom) the corresponding CCF between the simplified jet deviation and RM curves. In panel (a), the cross correlation at zero-lag is high, reflecting the alignment of the jet deviation and RM curves. The noughts and crosses denote the assumed magnetic field configuration, where the field is directed towards and away from the observer respectively. In panel (b), the zero-lag cross correlation is zero, which indicates in this case that the jet deviation and RM curves are out of phase. Here we assume that there is a uniform field component of the magnetic field along the LOS. The correlation at zero lag is used as a crude diagnostic between two classes of model for the Faraday-rotating region.}
    \label{fig:model_schematic}
\end{figure*}

To illustrate the expected observational signatures of each case, we introduce simplified toy models shown in Figure \ref{fig:model_schematic}, and use the CCF as a diagnostic of the relative phase behaviour between the jet deviation and RM structure. The models are not intended to capture the full complexity of jet dynamics or cluster turbulence, but to clarify how different physical origins map onto distinct CCF signatures.

\subsubsection{Jet-associated models}
\label{sec:jet_associated_models}

In a jet-associated scenario, the Faraday rotation arises from magnetised plasma physically connected to the jet or an associated sheath \citep[e.g.,][]{Kigure2004}. This may occur if magnetic fields are advected with the jet flow or amplified by instabilities that couple the magnetic field geometry to the oscillatory jet structure. A range of mechanisms—including current-driven kink instabilities, Kelvin--Helmholtz instabilities, magnetic tower flows, and sweeping magnetic twist models—have been proposed to produce quasi-periodic jet oscillations accompanied by coherent magnetic field distortions \citep{Nakamura2001, Kigure2004, Uchida2004, Nakamura2007, Mizuno2014, Singh2016, BarniolDuran2017, Mukherjee2020, Dong2020, Upreti2024}.

In this class of models, extrema in the RM curve are expected to coincide spatially with extrema in the jet deviation, as the magnetic field direction is coupled to the 3D jet orientation. The corresponding CCF therefore exhibits a strong peak (or trough, depending on magnetic field polarity) at zero lag, with additional peaks at integer multiples of the oscillation period. This behaviour is illustrated schematically in Figure~\ref{fig:model_schematic}, column (a), and provides a clear diagnostic of Faraday rotation that is co-spatial with, or directly influenced by, the jet.

\subsubsection{ICM-associated models}
\label{sec:icm_associated_models}

An alternative possibility is that the observed RM variations are generated primarily in the local ICM enclosed by the helices of the jet. In this scenario, the jet itself has little influence on the Faraday-rotating plasma, and the RM structure arises from systematic variations in LOS path length through the magnetised cluster medium along the jet’s winding trajectory. Corkscrew jets have previously been proposed as probes of ICM magnetisation under this assumption \citep[e.g.,][]{Johnston-Hollitt20151}, and jet--ICM interactions such as shocks may also imprint externally imposed RM structure \citep[e.g.,][]{Nolting2019}.

If RM variations are dominated by path-length effects in the ICM, extrema in the RM curve are expected to occur when the jet is maximally displaced toward or away from the observer. This introduces a systematic phase offset between the jet deviation and RM curves, resulting in a CCF that is close to zero at zero lag, with peaks appearing at non-zero lags corresponding to the phase offset and the sum of the phase offset and oscillation period. This behaviour is illustrated in Figure~\ref{fig:model_schematic}, column (b), and contrasts sharply with the jet-associated case.

\subsubsection{The Corkscrew Galaxy}

We now compare these toy model expectations with the observed CCF behaviour of the Corkscrew Galaxy. For the full jet, the CCF is weakly consistent with zero at zero lag (middle panel of Figure \ref{fig:cross_correlation}), with 72\% of the null ensemble exceeding the observed correlation in terms of amplitude. Taken in isolation, this would be more consistent with an ICM-dominated interpretation. However, as described in Section \ref{sec:correlations}, this interpretation is impacted by the fact that the Corkscrew Galaxy jet is not uniform along its length, exhibiting a clear morphological transition between eastern and western sections with different oscillation amplitudes and characteristic scales.

% the full-jet CCF also exhibits a sequence of non-zero peaks whose locations are inconsistent with the simple quarter-phase offsets predicted by the pure ICM path-length model. This discrepancy likely reflects the fact that 
We therefore consider the eastern and western jet sections separately (bottom of Figure \ref{fig:cross_correlation}, panels a) and b) respectively). In the eastern section, the CCF exhibits a relatively low null exceedance fraction at zero lag (7.6\%), indicating a degree of alignment may exist between the RM and jet deviations relative to the phase-randomised null curves. This behaviour is more consistent with a jet-associated Faraday-rotating region, where the magnetic field lines follow the jet path due to shear. In contrast, the western section shows a CCF consistent with zero at zero lag, suggesting the phase-shifted relationship between RM and jet deviation that would be expected from an ICM-dominated contribution.

Taken together, these results favour a mixed physical origin for the Faraday rotation along the Corkscrew Galaxy jet. In the eastern region, where the jet oscillates with smaller amplitude and higher frequency, RM variations are likely dominated by magnetised plasma associated with the jets or a sheath immediately surrounding the jets. Further downstream, as the oscillation amplitude increases and the jet samples a larger range of LOS depths, the contribution from the local ICM becomes increasingly important, producing the observed phase-shifted behaviour.

To assess the plausibility of this interpretation, we estimate characteristic magnetic field strengths using the standard RM relation,
\begin{equation}
\label{eq:RM}
\mathrm{RM} \; [\mathrm{rad\ m^{-2}}] = 0.812 \int_{0}^{L} n_e(s)\; [\mathrm{cm^{-3}}]\; B_\parallel(s)\; [\mu\mathrm{G}]\; \mathrm{d}s\; [\mathrm{pc}] \, ,
\end{equation}
where $n_e$ is the thermal electron density, $B_\parallel$ is the LOS magnetic field strength, and $L$ is the effective path length through the Faraday-rotating medium.

The ICM of the Norma cluster is detected in X-rays and exhibits an elongated morphology, with the X-ray centroid displaced by $\sim7.5$ arcmin to the north-west of the cluster centre, which is assumed to be at the position of the cD galaxy ESO 137-006 \citep{Jachym2014,Vollmer2024}. The Corkscrew Galaxy lies at a projected distance of $\sim15$--$20$ arcmin ($\sim300$--$400$ kpc) from this X-ray centre. Following Equation~15 of \citet{Vollmer2024}, and adopting a $\beta$-model ICM density profile with cluster parameters from \citet{Boehringer1996}, we estimate a thermal electron density of $n_e \simeq (6$--$9)\times10^{-4}\,\mathrm{cm^{-3}}$ at the location of the Corkscrew Galaxy.

In the western region, we assume that the RM variations arise from path-length differences through the ICM, with an effective LOS depth of $L \sim 80$ kpc based on the projected angular diameter of the corkscrew windings (Figure~\ref{fig:descriptive_maps_corkscrew}). For a tangled magnetic field with reversal scale $l$, the RMs add as a random walk, such that \citep{Gaensler2001}
\begin{equation}
B_\parallel \simeq \frac{\mathrm{RM}}{0.812\, n_e\, \sqrt{L l}} .
\end{equation}
Assuming large-scale magnetic coherence with $l \sim L$, we infer magnetic field strengths of $B \simeq 2$--$4~\mu$G in the western jet region. While slightly higher than values typically quoted for cluster outskirts \citep[e.g.,][]{Govoni2019,DominguezFernandez2019,Botteon2022}, many such estimates assume a steep radial decline in magnetic field strength following a simple $B \propto n_e^\eta$ model. Recent studies suggest that cluster magnetic fields may decline more slowly with radius, which could allow field strengths of a few microgauss to persist at large cluster-centric distances \citep[e.g.,][]{Osinga2025}. Reducing the reversal scale to $l \lesssim 10$ kpc increases the required field strengths to tens of $\mu$G, which are difficult to reconcile with typical ICM conditions in cluster peripheries. This implies that the Faraday-rotating ICM traced by the corkscrew morphology is characterised by magnetic fields coherent over scales of at least several tens of kiloparsecs (consistent with the overall smoothness observed in the RM curve in the top panel of Figure \ref{fig:curves_periodograms}), rather than a small-scale turbulent field. Additionally, a small-scale turbulent ICM field with coherence length $l \ll L$ would produce stochastic as opposed to periodic RM variations 
along the jet, as the random-walk contributions from many independent turbulent cells would destroy any coherent spatial patterns. The observed quasi-periodicity of the RM signal therefore independently requires large-scale field organisation on scales comparable to the jet winding scale, supporting the field strength argument against a purely small-scale turbulent ICM. 
% We also note that magnetic coherence on scales of tens of kiloparsecs does not necessarily imply detectable diffuse polarised synchrotron emission from the ICM. In cluster outskirts, the relativistic electron density 
% is expected to be low, and beam depolarisation and bandwidth depolarisation may strongly suppress polarised emission even in relatively ordered fields. The non-detection of diffuse ICM polarisation at current sensitivities is therefore not inconsistent with the large coherence lengths implied by our analysis.

For the jet-associated scenario, we assume a magnetised sheath with a characteristic thickness comparable to the observed jet width ($\sim20$ kpc; Figure~\ref{fig:descriptive_maps_corkscrew}). Given that the Corkscrew Galaxy is a tailed FRI source, we conservatively adopt the same thermal electron density as inferred for the local ICM. Under these assumptions, the observed RM amplitudes imply LOS magnetic field strengths of a few $\mu$G in the eastern region, consistent with values reported for other dynamically active radio galaxies \citep[e.g.,][]{Govoni2010,Guidetti2011,Anderson2018}.
Although these estimates rely on simplified geometries, they demonstrate that the mixed-origin interpretation is physically plausible and consistent with independent constraints. We therefore conclude that the Corkscrew Galaxy jet transitions from a jet-dominated Faraday screen near the host galaxy to increasing sensitivity to the magnetised ICM downstream.

\subsection{Methodological Insights for Future Studies of Corkscrew Jets} \label{sec:methodological_insights}

Our analysis of the Corkscrew Galaxy highlights several methodological requirements that are likely to be broadly applicable to studies seeking to detect and interpret quasi-periodic structure in radio jets and their associated RMs. We summarise here the key practical considerations that emerged and that should be taken into account in future work.

Resolving helical structure in Stokes~$I$ requires that the synthesised beam be a substantial fraction of the helix amplitude; without this, the projected oscillation cannot be reliably traced. Provided the helix is resolved, ridge-tracing algorithms such as \textsc{FilFinder} can recover a robust jet spine. These methods are most effective when the projected morphology resembles a smooth, approximately sinusoidal deviation rather than overlapping or self-intersecting loops, which can occur when the oscillation frequency is too high relative to the spatial resolution. The upcoming SKA-Mid's sub-arcsecond resolution at GHz frequencies will be critical in enabling the helix amplitude and winding scale to be resolved for a far larger population of sources than is currently accessible with ASKAP or MeerKAT. 
% SKA-Low may also further enable the detection of large-scale helical morphology in steep-spectrum, extended sources, where the characteristic spatial scales are sufficiently large.

The maximum recoverable oscillation period in either the jet deviation or RM signal is limited by the Rayleigh criterion, $P_{\mathrm{max}} = 1/(2R)$ (i.e. see Figure \ref{fig:curves_periodograms}). Periods exceeding this limit are not reliably distinguishable from long-term trends. Consequently, at least two well-resolved windings are required to claim periodicity, with a larger number strongly preferred for frequency-domain analyses such as Lomb--Scargle periodograms. Periods separated by less than $2R$ cannot be statistically distinguished and should be treated with caution. For sources that do not meet these criteria, alternative analyses may be more appropriate (see Section \ref{sec:corkscrew_jets_as_unique_astrophysical_lab}).

Reliable detection of RM periodicity also requires that the modulation amplitude exceeds the RM uncertainties by a factor of several. We expect that the SKA's broad fractional bandwidth will significantly reduce RM uncertainties, lowering the modulation amplitude required for reliable RM periodicity detection in the future. Our analysis also demonstrates the necessity of 
explicitly characterising the structure expected from correlated Galactic foreground RM, which can exhibit power across a broad range 
of spatial scales and is not well approximated by white noise. We therefore advocate the use of null tests tailored to the specific analysis method, such as comparing the observed RM periodogram to an ensemble of simulated foreground realisations with spatial correlation properties matched to the observed RM structure function of background sources in the field. The dense RM grids produced by SKA all-sky polarisation surveys will also provide the background source statistics needed to characterise Galactic foreground structure at the relevant angular scales for any individual source, making the foreground null tests described here straightforwardly applicable in the SKA era. In the context of cross-correlation analysis, such null tests are also essential for distinguishing genuine physical correlations from artefacts introduced by periodicity in only one of the input signals. 

Finally, where a jet exhibits clear changes in morphology—such as large variations in oscillation amplitude, frequency, or collimation—segment-wise analyses should be performed alongside any global treatment. This mitigates cancellation effects and allows for the possibility that different physical processes dominate the RM–jet relationship in different regions of the source.

\subsection{Corkscrew Jets as a Unique Astrophysical Laboratory in the SKA Era}
\label{sec:corkscrew_jets_as_unique_astrophysical_lab}
Helical radio galaxies provide an unusually rich laboratory for studying how AGN jets interact with their environments. However, our analysis of the Corkscrew Galaxy also highlights the observational and methodological limits of current data. Despite being the longest and brightest known helical system, the combination of sensitivity, resolution, and sampling are just sufficient to perform the frequency and spatial domain analyses presented in this paper. Sources with fewer than two resolved windings, or with substantially lower polarised SNR, will require alternative strategies. As we approach the SKA era, with precursors such as ASKAP and MeerKAT and ultimately the SKA, many more helical jet structures are likely to be discovered across a wide range of environments. These will enable a broader suite of analyses than is possible for a single source. We highlight several complementary directions:

\begin{enumerate}
    \item \textit{In-depth analyses of individual corkscrew jets.}
    The approach taken in this paper to establish whether jet morphology and associated RMs share a common spatial scale and phase relationship provides a template for future studies of well-resolved, high signal-to-noise helical sources. For such sources, analysis of jet geometry, RM periodicity and their relative phase can directly test the nature of the Faraday-rotating media and uncover whether the source can be used for constraining the magnetic field properties of the jet or its environment. In practice, this level of analysis will only be achievable for a subset of the brightest and most well-resolved corkscrew sources.

    \item \textit{In-depth analyses of specific regions of interest.}
    An alternative is to focus on particular regions in the jet–for example, individual jet bends, the region close to the core, or regions with interesting RM behaviour. High-resolution Stokes \emph{I} and polarisation imaging could allow detailed polarimetric analyses such as QU-fitting, intrinsic polarisation angle mapping, and depolarisation analysis. Depolarisation analysis could reveal more depolarisation in sections of the helix furthest from the observer, and polarisation angle mapping could reveal vectors that appear to follow the winding behaviour. Recent work by \citet{Sakemi2025}, who examined RM and polarisation structure in a localised bending region of a radio jet, illustrates the power of such focused studies to reveal how small-scale jet bending can couple with Faraday structure. The patchy polarisation observed in the Corkscrew Galaxy (Figure \ref{fig:descriptive_maps_corkscrew} panel b; also Figure \ref{fig:corkscrew_2dprojections}) suggests that depolarisation is occurring in at least some regions. However, the available in-band $\lambda^2$ coverage in the present observations is relatively limited and further fragmented by radio frequency interference removal within Band~2. Under these conditions, QU-fitting models become highly degenerate \citep[e.g.][]{Oberhelman2026}. Robustly distinguishing between internal and external Faraday-rotating structures generally requires broad, continuous $\lambda^2$ coverage and high angular resolution (e.g. \citealp{Sebokolodi2020}). Future broadband polarimetric observations with MeerKAT and the SKA will provide these capabilities and will therefore be better suited to disentangling the origin of the observed Faraday rotation and depolarisation.

    \item \textit{Statistical analyses of larger samples.}
    As the number of observed corkscrew jets increases, population-level analyses will become possible. Even when individual sources have too few windings for detailed individual analysis, ensembles of jets can be used to search for systematic trends in RM statistics. One valuable avenue would be to compare the RM distribution at jet winding extrema versus at the midpoints for many sources, searching for statistical differences in RM. This would be valuable for constraining how often jet-tracking RM structure occurs, and whether it is more commonly associated with jet-associated or local ICM Faraday screens. Large statistical samples could also allow correlation of RM behaviour with host galaxy or cluster properties.
    
    \item \textit{High-resolution 3D Magnetohydrodynamic (MHD)  Simulations.}
    All of the observational avenues above will be significantly strengthened by comparison with self-consistent MHD simulations of helical jets. Recent simulations of AGN jets already demonstrate how jet precession, instabilities and jet environment can generate complex helical morphologies and cocoon structures \citep[e.g.][]{Anjiri2014, BarniolDuran2017, Dong2020, Giri2022}.     
    Extending this work to explicitly simulate the evolution of helical or bent jets in realistic magnetised cluster environments, and to compute complimentary synthetic polarisation data would significantly compliment observational approaches. An example is the synthetic RM maps produced by \citet{Kigure2004}, and more recently, the synthetic synchrotron emission and polarisation maps produced by \citet{Jerrim2024, Upreti2024}. In particular, simulations that track both jet dynamics and Faraday-rotating plasma could be used to:
    \begin{itemize}
        \item predict how different Faraday-rotating media (magnetic fields associated with the jet, sheath, local ICM or foregrounds) map into observable RMs and phase relationships with the jet,
        \item quantify the impact of RM resolution and noise on being able to recover oscillatory structure,
        \item advance our understanding of the underlying formation mechanisms of corkscrew galaxies.
    \end{itemize}
\end{enumerate}

\section{Conclusions}
\label{sec:conclusions}

We have presented a broadband polarimetric study of the archetypal corkscrew radio galaxy ESO~137-G007 (the Corkscrew Galaxy) in the Norma cluster, based on ASKAP Band~2 (1296–1440 MHz) early science observations from the POlarisation Sky Survey of the Universe’s Magnetism (POSSUM). Our goal was to determine whether the quasi-periodic morphology of helical jet structure produces corresponding signatures in Faraday rotation, and whether such signatures can be used to constrain the three-dimensional structure of the jet and the physical origin of the magneto-ionic plasma responsible for the observed RMs.

By extracting one-dimensional representations of the jet trajectory and the RM signal along the jet, and analysing these using periodogram and cross-correlation techniques supported by tailored null tests, we find clear evidence for structured, non-random RM variations associated with the jet. The RM signal exhibits a dominant spatial scale consistent, within the Rayleigh resolution limit, with the lateral deviations of the corkscrew morphology. Foreground contributions from the Galaxy and from a purely foreground ICM screen are insufficient to account for the observed behaviour.

Our results indicate that the Faraday rotation associated with the Corkscrew Galaxy is not generated in a single, uniform screen. Instead, the data favour a mixed-origin scenario in which the dominant Faraday-rotating plasma changes along the jet. In the eastern, tightly-wound section of the jet, the RM variations are closely coupled to the jet morphology, consistent with a jet-associated or sheath-like Faraday screen. Further downstream, where the jet oscillations become larger and more loosely-wound, the RM behaviour becomes phase-shifted relative to the jet deviations, indicating increasing sensitivity to magnetised plasma in the local ICM sampled along the helical path.

Beyond these physical conclusions, this study highlights several methodological requirements for future corkscrew analyses. Reliable detection of quasi-periodic RM structure requires sufficient angular resolution to resolve multiple jet windings, RM modulation amplitudes that exceed measurement uncertainties, and analysis-specific null tests to distinguish genuine structure. Segment-wise analyses are essential when jet properties vary along their length, to avoid cancellation effects in global statistics.

While based on a single archetypal source, this work establishes a concrete observational framework for using radio galaxies with helical jet structures as probes of magnetised environments. The growing samples of helical jets being identified with ASKAP and MeerKAT, together with the sensitivity and resolution of the SKA, will enable population studies that test whether the jet–ICM transition observed here is common. Combined with targeted polarimetric analyses and complementary MHD simulations, such studies have the potential to turn helical radio galaxies into novel tracers of magnetic fields in galaxy cluster environments.

\section*{Acknowledgements}

C.~S.~A acknowledges funding from the Australian Research Council in the form of FT240100498. POSSUM is partially funded by the Australian Government through an Australian Research Council Australian Laureate Fellowship (project number FL210100039 awarded to NMM-G). SPO acknowledges support from the Comunidad de Madrid Atracción de Talento program via grant 2022-T1/TIC-23797, and grant PID2023-146372OB-I00 funded by MICIU/AEI/10.13039/501100011033 and by ERDF, EU. H.~S acknowledges support from JSPS KAKENHI Grant Numbers JP22K20386, JP23K13148, and JP26K17195. 

This scientific work uses data obtained from Inyarrimanha Ilgari Bundara / the Murchison Radio-astronomy Observatory. We acknowledge the Wajarri Yamaji People as the Traditional Owners and native title holders of the Observatory site. The Australian SKA Pathfinder is part of the Australia Telescope National Facility (https://ror.org/05qajvd42) which is managed by CSIRO. Operation of ASKAP is funded by the Australian Government with support from the National Collaborative Research Infrastructure Strategy. ASKAP uses the resources of the Pawsey Supercomputing Centre. Establishment of ASKAP, the Murchison Radio-astronomy Observatory and the Pawsey Supercomputing Centre are initiatives of the Australian Government, with support from the Government of Western Australia and the Science and Industry Endowment Fund. The POSSUM project (https://possum-survey.org) has been made possible through funding from the Australian Research Council, the Natural Sciences and Engineering Research Council of Canada, the Canada Research Chairs Program, and the Canada Foundation for Innovation.

%%%%%%%%%%%%%%%%%%%%%%%%%%%%%%%%%%%%%%%%%%%%%%%%%%
\section*{Data Availability}
The visibility data used in this paper are publicly available on CASDA \citep{Huynh2020}. Additional data products are available from the authors upon reasonable request.


\bibliographystyle{mnras}
\bibliography{bexample.bib}

%%%%%%%%%%%%%%%%%%%%%%%%%%%%%%%%%%%%%%%%%%%%%%%%%%

%%%%%%%%%%%%%%%%% APPENDICES %%%%%%%%%%%%%%%%%%%%%

\appendix

\section{Details of methods used}
\label{sec:methodsappendix}
In Section \ref{sec:resultsandanalysis}, we outlined the main methods used in our analysis. In this appendix we provide the full methodological details of those procedures, including RM reliability criteria and Galactic foreground correction (Section \ref{sec:rm_reliability}), the pseudo‑3D visualisation technique used for qualitative RM–morphology comparison (Section \ref{sec:pseudo3D_polcubes}), construction of the jet deviation and RM curves (Section \ref{sec:curves}), the Lomb–Scargle periodogram analysis and null test (Section \ref{sec:periodogram}), and the CCF and significance test (Section \ref{sec:ccf}).

\subsection{RM Reliability and Galactic Foreground Correction}\label{sec:rm_reliability}
Reliable RM measurements from RM synthesis at POSSUM frequencies require a polarised $\text{SNR}\gtrapprox7$ to avoid significant Faraday depth bias and spurious detections \citep{Macquart2012, George2012,Hales2012,Thomson2023}. Following previous work, we apply a polarised $\text{SNR}\geq7$ threshold to the \emph{PI} and RM maps prior to any further analysis. We calculated the RM uncertainties ($\delta$RM) using:
\begin{equation}
\delta \text{RM} = \frac{W}{2\text{SNR}},
\end{equation}
where $W$ is the FWHM of the main peak of the RM spread function (RMSF) \citep{Brentjens_deBruyn_2005}, equal to $340~\radms$ for POSSUM Band 2 (Table~\ref{table:observational_details}). To remove the RM contribution from Galactic foreground Faraday rotation, we assume the foreground varies smoothly on the angular scale of the source and apply a constant correction by subtracting the median RM of all pixels passing the SNR cut ($-19~\radms$). All RMs quoted and analysed in this paper are referenced to the foreground-subtracted values.

To quantify residual small-scale foreground structure, we compute the second-order RM structure function of the surrounding POSSUM RM grid, defined as
\begin{equation}
D(\theta)
=
\left\langle
\left[ \mathrm{RM}(\mathbf{x}) - \mathrm{RM}(\mathbf{x}+\boldsymbol{\theta}) \right]^2
\right\rangle,
\end{equation}
where $\mathbf{x}$ is the sky position vector and the angle brackets denote an average over all independent source pairs separated by angular distance $\theta$. We restrict the sample to sources within $\pm1^{\circ}$ in Galactic latitude and $\pm10^{\circ}$ in Galactic longitude of the Corkscrew Galaxy, divide pairs into 30 logarithmically spaced bins spanning $0.01$--$10^{\circ}$ (discarding bins with fewer than ten pairs), and compute the mean squared RM difference per bin. We subtract the contribution from measurement noise by removing the pairwise uncertainty term from each RM difference. The noise contribution to the structure function for a pair of sources $i$ and $j$ is
\begin{equation}
D_{\mathrm{noise},ij} = \sigma_{\mathrm{RM},i}^{2} + \sigma_{\mathrm{RM},j}^{2},
\end{equation}
where $\sigma_{\mathrm{RM},i}$ and $\sigma_{\mathrm{RM},j}$ are the uncertainties of the individual RM measurements. Vertical dashed lines in Figure~\ref{fig:structurefunc} indicate the degree-scale Galactic coherence length (black) and the approximate jet winding scale (red) and error bars represent the standard error of the mean within each separation bin.

\subsection{Pseudo-3D Visualisation}\label{sec:pseudo3D_polcubes}

Following the approach introduced by \citet{Rudnick2024}\footnote{\url{https://github.com/candersoncsiro/rmsynth3d}}, we employed a pseudo-3D visualisation technique to represent Faraday structure in the (RA, Dec, RM) space. This method constructs polarisation cubes by combining the peak \emph{PI} $ P(\text{RA}, \text{Dec}) $ and its associated Faraday depth $ \Phi(\text{RA}, \text{Dec}) $.

In these cubes, the \emph{PI} at each sky position is mapped to a corresponding pixel along the Faraday depth axis, with a degree of smoothing applied that preserves flux. This method provides a simplified representation of the full FDF, which has the benefits of reducing noise, highlighting dominant Faraday emission features, and mitigating spurious sidelobe artefacts in the resulting data cube.

By analysing the spatial correlations between \emph{PI} structures and variations in RM through various projections and dynamic visualisations (e.g., animations cycling through projections), this technique enables the differentiation of physical origins for polarisation features, such as stochastic foreground Faraday screens versus intrinsic magnetised plasma within the source \citep{Rudnick2024}.

Recognising the potential of this method for visually identifying spatial correlations between the corkscrewing jet features and RM variations, we applied it to the Corkscrew Galaxy. We generated projected 2D maps in both plane-of-sky coordinates (RA, Dec.) and Faraday depth space (RM, RA), as shown in Figure \ref{fig:corkscrew_2dprojections}.

\subsection{Quantifying Jet Morphology and Faraday Rotation Structure}
\label{sec:curves}

To examine the relationship between the Corkscrew jet oscillations and
RM structure quantitatively, we constructed two diagnostic curves: jet deviation and RM. The jet deviation curve quantifies lateral displacements of the jet relative to a defined jet axis, characterising its morphology in the plane of the sky. The RM curve traces variations in RM along the jet, offering insight into the magnetoionic structure along the LOS. The following
subsections detail the construction of each curve.

\subsubsection{Jet Deviation Curve: Measuring Lateral Displacements}
\label{sec:jet_deviation}

We used the Python package \textsc{FilFinder}\footnote{\url{https://fil-finder.readthedocs.io/en/latest/}} \citep{Koch2015} to extract the jet deviation curve. {\sc{FilFinder}} was designed for extraction and analysis of filamentary structure in molecular clouds, but as we demonstrate, its methods can effectively trace the morphological features of radio galaxies. Following the \textsc{FilFinder} tutorial, we flattened the Stokes \emph{I} MFS image prior to creating the filament mask. There are several parameters that set the masking behaviour. We specified values for `size\_thresh' and `glob\_thresh' such that the resulting mask best captured the jet morphology. The masks were reduced to single-pixel-width skeletons, which were pruned based on average branch intensity and length to remove insignificant spurs. To improve spatial resolution, the resulting skeletons were smoothed using a locally weighted scatterplot smoothing (LOWESS) approach \citep{Cleveland1979} as implemented in the \textsc{statsmodels} Python library, with the smoothing parameter chosen to balance smoothness and fidelity to the original data.

A representative jet axis line was defined to characterise the general jet orientation, about which lateral displacements (jet deviations) were measured as perpendicular distances from the smoothed skeleton points to this axis. As the Corkscrew Galaxy exhibits a clear orientation change at RA, Dec (J2000), $\sim16\text{h}13.5\text{m}$, $-60\degr34\arcmin$, we defined two jet axis lines. Variations in axis positioning were tested and found not to significantly impact the derived jet deviations. We also truncate the jet deviation curve to the portion of the jet with clear oscillations (i.e, we remove a small portion of the eastern-most side of the jet, as the oscillations are poorly resolved and difficult to characterise using \textsc{FilFinder}). We also calculate the cumulative angular distance along the jet axis lines (`distance along jet axis') from the start of the jet to each point’s projection onto the jet axis.

To estimate uncertainties on the jet deviation curves, we employ a Monte Carlo approach. We generated 100 random realisations of the jet axis by perturbing the jet axis line positions with uniform random offsets of up to $\pm5$ pixels at each end. Additionally, positional uncertainties derived from the Stokes \emph{I} maps were incorporated as amplitude shifts. For each realisation, the jet deviation curve was recalculated, producing a sample from which error estimates were derived. This accounts for systematic uncertainties in axis definition and intrinsic positional errors in the Stokes \emph{I} MFS images.

To standardise the spatial sampling of jet deviation curves, the unevenly spaced `distance along the jet axis' points were interpolated onto a uniform grid of 2000 points spanning the full range of measured distances along the jet axis lines. Cubic spline interpolation was employed for both the jet deviation and their associated uncertainties. The sample of jet deviation curves generated using the Monte Carlo approach described above were similarly interpolated onto the uniform grid using cubic splines and used for measurement uncertainty in subsequent analysis.

\subsubsection{RM Curves: Mapping Faraday Rotation Along The Jets}
\label{subsec:rms}

To map RM variations along the jet, we extracted RM values within a sequence of `strips' centred at each point along along the smoothed jet trajectory. At each point, a rectangular strip with fixed width (1 pixel) and fixed length (70 pixels, chosen such that the full width of the jet was encapsulated at each point along its extent) was constructed. Each rectangular strip was centred on the corresponding skeleton coordinate and rotated so that its width was along the jet axis, and its length was perpendicular to the jet axis.

In each strip, we selected all RM values (with the \emph{PI} threshold and foreground subtraction applied, as described in Section \ref{sec:rm_reliability}), restricted to regions inside the source mask. To further mitigate the influence of outliers, we computed a 5\% trimmed mean for the RM distribution within each strip, which was adopted as the characteristic RM value at that position along the jet.

Measurement uncertainties were assessed via a Monte Carlo approach. For each strip, $1000$ synthetic RM distributions were generated by sampling each pixel’s RM from a normal distribution with a mean set by its measured RM and a standard deviation equal to its uncertainty (dRRM). For each realisation, the trimmed mean was recomputed, and the final uncertainty was determined as the standard deviation of these simulated trimmed mean RMs.

The final RM curves were detrended via linear regression to remove large-scale linear gradients along the jet axis. Standard errors for the slope and intercept were computed under the assumption of normally distributed residuals, providing estimates of the uncertainty in the linear trend. For subsequent uncertainty analysis, we randomly sampled slopes and intercepts from within the standard errors from the linear regression and subtracted the corresponding fitted lines from the RM measurements to produce 100 realisations of the RM curves for error estimates in the subsequent analysis. Additionally, RM uncertainties were incorporated as amplitude shifts.

Cubic spline interpolation was employed for both the RM curves and the realisations to standardise the spatial sampling of the RM curve along the re-sampled `distance along the jet axis' points (described in Section \ref{sec:jet_deviation}).

\subsection{Periodogram Analysis}
\label{sec:periodogram}
We characterised the spatial variability in the jet deviation and RM curves using Lomb–Scargle periodograms. We converted spatial frequency along the jet axis to spatial scale by taking the inverse of frequency, and identified dominant scales as peaks in the spectra. For visual comparison, we plot the jet deviation and RM curves with their periodograms, with vertical lines marking the dominant spatial scales obtained from the peak of each curve's periodogram spectrum.  We also computed the periodograms of each of the 100 realisations of the jet deviation and RM curves, using the standard deviations of the resulting periodograms as the error estimate for the true jet deviation and RM periodograms.

To evaluate statistical significance, we constructed a null distribution of 1000 simulated Galactic foreground RM profiles using a Gaussian random field (GRF) approach. The spatial covariance structure of the simulated fields was derived from the observed RM structure function of background sources within $\pm1^{\circ}$ in Galactic latitude and $\pm10^{\circ}$ in Galactic longitude of the Corkscrew Galaxy (Section~\ref{sec:GalFG}), which is well described by a power-law model $D(\theta) \propto \theta^{\alpha}$ with $\alpha = 0.48$, saturating at a correlation length of $l_0 \approx 200$ arcmin.

Each GRF realisation was generated on a $1145 \times 1145$ pixel grid at 2~arcsec~pixel$^{-1}$, matching the spatial extent and resolution of the observed RM map, by Fourier-transforming a white noise field weighted by the square root of the analytically derived power spectrum corresponding 
to a fitted covariance model:

\begin{equation}
    C(r) = \sigma_{\rm rm}^2 \exp\left[-(r/l_0)^{\alpha}\right],
\end{equation}

\noindent where $\sigma_{\rm rm}$ is the RM standard deviation derived from the structure function amplitude, and $r$ is the pixel separation in arcminutes. Each realisation was normalised to the expected standard deviation within the finite field. A 5\% trimmed-mean profile was then extracted from a 70-pixel-wide strip at that position, in an identical manner to the extraction of the observed RM profile 
(Section~\ref{filfindermethodandresults}; \ref{subsec:rms}). The Lomb--Scargle periodogram of each 
simulated profile was computed and normalised to the peak power of the observed RM periodogram, forming the null distribution against which the 
observed RM peaks are assessed. The strength of any peak is quoted as a multiple of the 84th percentile of the null distribution at that spatial scale.

We follow the recommendations outlined by \citet{RamirezDelgado2025} regarding the use of the Rayleigh criterion to define minimum and maximum detectable periods and avoid false positive detections caused by uneven sampling or sampling of insufficient length. The Rayleigh frequency of the full jet dataset is
\[
R = \frac{1}{T} = 0.00086 \,\mathrm{1/arcsec}, 
\]
where $T$ is the total spatial extent of the curves in arcseconds, giving a minimum resolvable oscillation of $2R$ and a corresponding maximum detectable period of
\[
P_{\max} = \frac{1}{2R} = 581.6\,\mathrm{arcsec}.
\]
As the Rayleigh frequency corresponds to the width of the narrowest feature that can be
distinguished in the periodogram, it also provides an estimate of
the frequency uncertainty for a given peak in the periodogram. We convert the resolution-limited frequency uncertainty into an uncertainty in the period using standard error propagation, with final uncertainties quoted according to:
\begin{equation}
    \sigma_P 
    \;=\;
    \frac{1}{T f^{2}}
    \;=\;
    \frac{P^{2}}{T},
\end{equation}
where $f$ is the frequency of the relevant peak and P is the corresponding period in spatial scale (arcseconds).

\subsection{Cross-correlation Function Analysis}\label{sec:ccf}
We compute the CCF between the jet deviation and RM curves using the \textsc{Spectrum} 
Python package \citep{Cokelaer2017}, after mean-centring both curves to remove the 
effect of any constant offset. Uncertainties are estimated via 100 Monte Carlo 
realisations of both curves, with the standard deviation of the resulting ensemble 
adopted as the CCF uncertainty. Lag indices are converted to angular distances along 
the jet axis.

To construct the null distribution described in Section~\ref{sec:correlations}, we 
generate 500 synthetic RM curves using the Timmer--K\"{o}nig method 
\citep{TimmerKoenig1995}, which randomises the phases of the original RM signal while 
preserving its power spectrum. Each synthetic curve is cross-correlated with the 
observed jet deviation curve. The strength of a CCF feature at a given lag is 
characterised by the fraction of null CCFs that exceed the observed amplitude at that 
lag. Prominent peaks are identified using a minimum prominence threshold of 300, 
corresponding to the $\sim$95th percentile of the null prominence distribution, which 
filters small fluctuations and ensures only globally unusual features are retained. 
Peak prominence is defined as the vertical distance between the peak and the lowest 
contour line encircling it but containing no higher peak.

The interpretable lag range is restricted to ensure CCF features are physically 
meaningful. For two signals sharing a dominant oscillatory scale $P$, the CCF 
exhibits quasi-periodic structure with successive same-sign peaks separated by 
approximately $P$. A relative phase shift of $P/2$ moves the signals from maximum 
alignment to maximum anti-alignment, defining a minimum number of overlapping points 
required for a feature to be meaningful. We therefore require at least $\Delta\tau = 
P/2$ overlapping points at all plotted lags, ensuring we interpret only features that 
could plausibly be produced by the periodogram-derived scale $P$. The smallest 
resolvable lag step is set by the spatial sampling interval $\Delta x \simeq 
0.6~\mathrm{arcsec}$, so CCF peak positions cannot be determined more precisely than 
this.

%%%%%%%%%%%%%%%%%%%%%%%%%%%%%%%%%%%%%%%%%%%%%%%%%%

% Don't change these lines
\bsp	% typesetting comment
\label{lastpage}
\end{document}